\documentclass[fleqn,10pt]{wlscirep}
\usepackage[utf8]{inputenc}
\usepackage[T1]{fontenc}
\title{A new advance on dimensional-aware scalar, vector and matrix operations in C++}

\author[1,2]{Eduard George Stan}
\author[1]{Dan Andrei Ciubotaru}
\author[1]{Michele Renda}
\author[1*]{C\u{a}lin Alexa}
\affil[1]{IFIN-HH, Horia Hulubei National Institute for R\&D in Physics and Nuclear Engineering, Particles Physics Department, M\u{a}gurele, Romania}
\affil[2]{University of Bucharest, Faculty of Physics, M\u{a}gurele, Romania}

\affil[*]{\textit{Corresponding author:} calin.alexa@nipne.ro}

\usepackage[frozencache]{minted}
\setminted{fontsize=\footnotesize,frame=lines,breaklines=true}
\usepackage{booktabs}
\usepackage{cleveref}
\usepackage{siunitx}
\usepackage{newtxmath} 
\usepackage{amsmath}
\usepackage{amssymb,wasysym}
\usepackage{url}
\usepackage{hyperref}
\usepackage{todonotes}
\usepackage{feynmp-auto}
\usepackage{subfigure}
\usepackage{pdflscape}

\DeclareUnicodeCharacter{2080}{$_0$}
\DeclareUnicodeCharacter{0393}{$\Gamma$}
\DeclareUnicodeCharacter{03C4}{$\tau$}
\DeclareUnicodeCharacter{2081}{$_1$}
\DeclareUnicodeCharacter{2082}{$_2$}

\begin{abstract}
	We review the dimensional check problem of the high-level programming languages, discuss the existing solutions, and come up with a new solution suited for scientific and engineering computations. Then, we introduce \texttt{Univec}, our \texttt{C++} library designed to make scalar, vector, and matrix operations using units of measurement. Moreover, \texttt{Univec} supports dimensional-aware operations for complex numbers, quaternions, octonions, and sedenions. We provide tables of the relevant functions and operators implemented. Our library was compared with several existing solutions, and the results are shown in the performance section. Finally, we present our future plans for improving the current implementation.
\end{abstract}
\begin{document}

\flushbottom
\maketitle

\thispagestyle{empty}

\graphicspath{ {./images/} }

\section{Introduction}
Dimensional analysis is one of the most powerful tools available in physics to verify the correctness of mathematical formulas describing physical processes. However, this tool alone is not able to spot all possible mistakes. It allows only a fast check, highlights faulty operations, and provides a consistent examination of the whole process.

Nowadays, complex simulations and derivations are performed using computing algorithms taking advantage of high-speed calculation and numerical correctness due to the advances in computing technology. However, when using complex algorithms spanning thousands of source code lines divided into tens of ﬁles, we tend not to automatically trust the results or, much worse, accept them blindly. Moreover, because there is no bug-free software implementation, we have to mitigate the impact of possible errors using countless checks in code and proper testing procedures, which are often as time-consuming as the development itself.

Therefore, an appropriate approach is to use development techniques that could help reduce the overall cost of software development \cite{mayerhofer_adding_2016}.
Unfortunately, due to historical reasons, most high-level programming languages lack a dimensional check as a language feature; they work on floats and integers, leaving to the programmer's judgement the responsibility to keep track of their semantic meaning, leading to many potential mistakes.

Several libraries provide dimensional analysis check at compile time or at run-time. However, as shown by Mc.Keever \cite{mckeever_unit_2021}, there is an evident reticence to adopt such solutions due to the following reasons:
\begin{itemize}
	\item Difficulty in using unfamiliar and complex libraries
	\item Fear of adding external dependencies
	\item Concerns about the performance impact
	\item Possible limitations on using external libraries or complex data structures such as matrices and vectors.
\end{itemize}

In \cref{sec:existing_solution}, we review the existing solutions available in the most common programming languages, we describe our proposed solution and present concrete examples in \cref{sec:our_implementation}, we discuss the testing approach in \cref{sec:testing} and present a performance analysis in \cref{sec:performance}. Limitations of the current development and plans for improvements are discussed in \cref{sec:limits} and conclusions are drawn in  \cref{conclusions}.

\section{Existing solutions} \label{sec:existing_solution}
A comprehensive review of several existing solutions for dimensional analysis in programming languages can be found in Preussner's work \cite{preussner_dimensional_2018}.
Apple's Swift \cite{apple_swift_2022} and \texttt{F\#} \cite{kennedy_types_2010} are two widespread programming languages that provide native support for UoM-aware (units of measure) quantities.
Promoting UoM analysis to a language feature has the benefit of immediate access to meaningful error messages. Unfortunately, only a few programming languages offer native support for this important topic. Instead, most languages have an external library that relies on generic programming techniques.

The most popular solution for \texttt{C++} is \texttt{Boost.Units} \cite{boost_units}, which makes possible the UoM validation at compile time. Its main advantage is that it is included in the \texttt{Boost} framework, a comprehensive set of libraries that are used as a launchpad for features that are later included in the language's standard library. Unfortunately, \texttt{Boost} is quite often seen as a heavy dependency.
For the \texttt{FORTRAN} language there is the \texttt{PHYSUNITS} module \cite{petty_automated_2001}
which provides support for dimensional-aware routine creation.
While this solution is versatile, allowing working with both \texttt{F77} and \texttt{F90} code,
because the dimension information is carried alongside numerical values,
\texttt{PHYSUNITS} introduces a run-time execution overhead in terms of speed and memory usage.
Dimensional analysis is not limited to compiled languages, \texttt{Python} supports run-time dimensional analysis checks
thanks to libraries like \texttt{Pint} \cite{grecco_pint_2022}.

Almost all existing libraries do not provide native support for vectorial computation.
However, for vectorial operations in high-energy physics there are two solutions:
\texttt{CLHEP} \cite{lonnblad_clhepproject_1994} and \texttt{Eigen} \cite{eigenweb}.
They provide representations for both 2D and 3D vectorial quantities and n-dimensional matrices, but none of them has built-in support for dimensional-aware units, relying only on primitive numeric types available in \texttt{C++}.

We have to mention that our approach is not the first attempt to solve this specific issue. For example, the \texttt{CORSIKA 8} simulation tool \cite{dembinski_corsika_2019}, a \texttt{C++} rewrite of \texttt{CORSIKA} \cite{heck_corsika_1998}, uses a similar solution to address the same problem of combining unit systems with vector arithmetic. The approach adopted there was to combine \texttt{Eigen v3} \cite{eigenweb} with the \texttt{PhysUnits} library \cite{physunitsweb}, allowing them to perform dimensional-aware vector arithmetic without any performance penalty \cite{dembinski_corsika_2019}.

\section{Our Implementation} \label{sec:our_implementation}
To improve the accuracy and lower the run-time overhead of the scientific and engineering computations, we developed \texttt{Univec}, our \texttt{C++} library that allows scalar, vector, matrix, complex, quaternion,  octonions, and sedenion operations integrating the \texttt{Boost.Units}\cite{boost_units} library features.

A summary of the functions and aliases available for vector and matrix types can be found in \cref{tab:common_operations,tab:nions_methods,tab:vector_operators,tab:matrix_methods,tab:classes-frame,tab:classes_and_aliases,tab:matrix_is_methods,tab:common_operations}, while the complete source code, released under \texttt{LGPLv3} license, can be found at the address \url{https://gitlab.com/micrenda/univec}

All the methods implemented return a copy of the original vector. However, for methods for which the result is compatible with the calling class type, we created an in-place version of the method, which modifies the object instance. These methods start with the prefix \texttt{do} (e.g. \texttt{doConj()}) to be easily differentiated from the copy version of the same method. These methods are usually faster than those that copy data because they do not allocate additional memory for their operations.

Our implementation heavily uses generic template programming, shifting the cost of the dimensional check at compiling time and removing, in principle, any run-time overhead.
However, the drawback is an increased usage complexity, which may discourage non-experienced \texttt{C++} developers. We tried to mitigate this issue by providing custom descriptive error messages and a set of default quantity aliases covering the most common quantities used in physics. This approach allows easy-to-remember type names when defining a variable, such as:
\begin{minted}{c++}
dfpe::QtySiVelocity my_var;
\end{minted}
which is a much shorter name than its canonical name:
\begin{minted}{c++}
boost::units::quantity<boost::units::si::velocity>  my_var;
\end{minted}

We decided to keep the set of aliases in SI, GCS and Gauss units, isolated from the main library development and were published as a separated project, at the address \url{https://gitlab.com/micrenda/qtydef}

\subsection{Vector operations} \label{sec:vector}
\texttt{Univec} allows writing a compact and semantically clear code with dimensional-check at compile-time. This is done using a set of classes, shown in \cref{tab:classes_and_aliases}, representing different types of quantities, such as vectors and matrices. In our implementation we tried to prioritize clarity and performance, two objectives that are difficult to reconcile because fast code tends to be quite complex and difficult to read. We implemented all the $n$-dimensional Cartesian vector classes, including complex numbers and quaternions, in a single header-only class. This approach required heavy use of template programming. However, it provides two important benefits: code duplication was significantly reduced, and we got a highly optimized implementation because the method resolution is performed at compile-time since we do not have inheritance and virtual functions, allowing efficient compiler optimizations. Furthermore, the complexity of the code due to generic programming was mitigated using the last \texttt{C++} keywords like \texttt{requires}, which improve substantially the code readability.

During the design, we realized that complex numbers, quaternions, octonions, and sedenions methods were a super-set of the methods available in Cartesian $n-$dimensional vectors (see \cref{tab:nions_methods}). Therefore, we used the same classes for these entities, marking complex quantities as vectors with a negative dimensional number. In this way, we can represent complex numbers as \texttt{VectorC<-2>}, quaternions as \texttt{VectorC<-4>}, etc., enabling and disabling some methods using the $N$ template parameter. However, for better readability, a set of alias for common entities is provided, as shown in \cref{tab:classes_and_aliases}. This solution allowed a high degree of code reuse and painless conversion between the entities without requiring inheritance and virtual method resolutions at runtime.

\subsection{Matrix operations} \label{sec:matrix}
During the development of the vectorial operations, we realized that matrix operations would be needed soon, which shall also beneﬁt from the same UoM validation. The main difficulty was that not all operations were available for any matrices. For example, an operation like the determinant of a matrix is deﬁned only for a square matrix. For this reason, we made heavy use of the \texttt{C++} concept \texttt{requires}. Although this severely limits its usage only within projects that use at least \texttt{C++20}, it allows a clear definition of the function usage constraints, and produces meaningful error messages when these conditions are not met.

As a design decision, matrix dimensions can be set using integer templates, as shown in \cref{tab:classes_and_aliases}: this means that matrix dimension can not be changed at runtime, allowing non-negligible performance optimizations by the compiler. As we can observe in \cref{tab:matrix_operations,tab:matrix_is_methods,tab:matrix_methods}, we decided to limit the method's implementations to the one which have a deterministic and non-iterative solution, omitting methods requiring matrices decomposition. In a future release, we may add support for this class of methods, allowing the calculation of eigenvalues, eigenvectors, matrix $L_2$ norms and similar quantities.

\subsection{Floating point limitations and tolerances}

Representation of floating-point number is a tricky point in any software package that handle non-integer quantities.  The IEEE Standard for Floating-Point Arithmetic (\texttt{IEEE 754}) \cite{ieee_floating_2019}, is the de facto standard in modern computers, and \texttt{C++} supports it using the \texttt{float} and \texttt{double} variable types. This format allows storing any number inside a fixed size mantissa and exponent, allowing the representation of up to \num{1e308} or down to \num{1e-308}, with around \num{15}-digits precision.

One first issue is that some rational numbers that are perfectly representable on a base-\num{10} numeric system can not be expressed on a base-\num{2} numeric system and inevitably introduce a rounding error.

Another issue is that trigonometric and transcendental functions in \texttt{C++} are calculated using a series sum and are truncated after a specific precision is achieved. This is usually not a problem in the engineering and scientific fields because we are interested in the approximate values of our calculation, given that we do not fall into known floating-point pitfalls. However, our library is quite sensitive about this issue because often, we have to check if two vectors are parallel or perpendicular, or if a matrix is diagonal: these actions do require to check the result against the exact values, which is tricky when working with floating-point quantities.

To manage this last issue, we introduced the concept of tolerance, represented by the class \texttt{Tolerance}. This class can be passed as a template parameter to any function that requires comparing two values for equality (usually, but not always, the methods starting with the "is" prefix). We provide a default implementation which compares two values as equals if they equal to the \num{10}th decimal digit. While this approach is not quite robust, it has the advantage of being relatively fast and easy to understand. If a different approach is required, the user can provide a custom class for all the methods or just for some specific calls.

\begin{table}
	\centering
	\begin{tabular}{p{.13\linewidth} p{.30\linewidth}  p{.48\linewidth} r}\toprule
		Operation                                                                   & Method                            && Note\\
		\midrule
		$\textbf{v} \cdot \textbf{u}$                                               & \texttt{v.dot()}                  & Dot product & \\
		$\textbf{v} \times \textbf{u}$                                              & \texttt{v.cross(u)}               & Cross product& \\
		$\varangle(\textbf{v} , \textbf{u})$                                        & \texttt{v.angle(u)}               & Angle between two vectors& \\
		$\hat{\textbf{v}}$                                                          & \texttt{v.versor()}               & Versor associated & $\ast$\\
		$\hat{\textbf{v}}$															& \texttt{v.versorDl()}             & Versor associated (dimensionless)& \\
		$\pm |\textbf{v}|/|\textbf{u}| \text{ if } \textbf{v}\parallel\textbf{u}$   & \texttt{v.scale(u)}               & Ratio between parallel vectors & \\
																					& \texttt{v.rotate(a,ax)}           & Rotate around z-axis ($2D$) or an arbitrary axis ($3D$) & $\ast$\\
		$\textbf{v} \oplus \textbf{u}$												& \texttt{v.directSum(u)}           & Direct sum & \\
		$\textbf{v} \odot \textbf{u}$												& \texttt{v.elementwiseProduct()}   & Element-wise product  & \\
		$\textbf{v} \oslash \textbf{u}$												& \texttt{v.elementwiseDivision()}  & Element-wise division & \\
		\midrule
		$|\textbf{v}|$                                                              & \texttt{v.norm()}                 & Euclidean norm & \\
		$|\textbf{v}|_{1}$                                                          & \texttt{v.normL1()}               & $L_1$ norm & \\
		$|\textbf{v}|_{p}$                                                          & \texttt{v.normL(p)}               & $p$ norm   & \\
		$|\textbf{v}|_{\infty}$                                                     & \texttt{v.normLInf()}             & Infinity norm & \\
		$|\textbf{v}|^2$                                                            & \texttt{v.normSquared()}          & Euclidean norm squared & \\
		\midrule
		$\textbf{v} \parallel \textbf{u}$                                           & \texttt{v.isParallel(u)}          & Are parallel & \\
		$\textbf{v} \perp \textbf{u}$                                               & \texttt{v.isPerpendicular(u)}     & Are perpendicular & \\
		$\textbf{v} \uparrow \uparrow \textbf{u}$                                   & \texttt{v.isSameDirection(u)}     & Are parallel and have same directions & \\
		$\textbf{v} \uparrow \downarrow \textbf{u}$                                 & \texttt{v.isOppositeDirection(u)} & Are parallel and have opposite directions & \\
		\midrule
		$ \textbf{v} = \textbf{u}$													& \texttt{v.isNear(u)}              & Are the same & \\
		$ \textbf{v} = 0$															& \texttt{v.isNull()}               & Is zero & \\
																					& \texttt{v.isNan()}                & Has any NaN component & \\
																					& \texttt{v.isNormal()}             & Are all components neither infinity, NaN, zero or subnormal & \\
		$ \textbf{v} \ne \pm \infty$												& \texttt{v.isFinite()}             & Are all components finite & \\
		$ \textbf{v} = \pm \infty$													& \texttt{v.isInfinite()}           & Has any infinite component & \\
		\bottomrule
	\end{tabular}
	\caption{\noindent List of functions implemented for each type of vector. Cross product is only implemented for $3D$ vectors, while the angle is implemented only for $2D$ and $3D$ vectors.  The $\ast$ marks the disponibility of an extra operation, which starts with \texttt{do} prefix (e.g. \texttt{doVersor}), that works inplace on the vector.}
	\label{tab:common_operations}
\end{table}

\begin{table}
	\centering
	\begin{tabular}{p{.20\linewidth} p{.23\linewidth}  p{.48\linewidth} r}\toprule
		Operation                                                  & Method                 & Description                                & Note\\
		\midrule
		$ \mathbf{q}^*$                                            & \texttt{q.conj()}      & Conjugate                                  & $\ast$\\
		$\mathbf{q}^{-1}$                                          & \texttt{q.inv()}       & Reciprocal                                 & \\
		$0 + b \mathbf{i} +  c \mathbf{j} + d \mathbf{k} + \ldots$ & \texttt{q.pure()}      & Vector complex, quaternion, octonion, etc. & \\
		$a + 0 \mathbf{i} +  0 \mathbf{j} + 0 \mathbf{k} + \ldots$ & \texttt{q.scalar()}    & Scalar complex, quaternion, octonion, etc. & \\
		$(b, c, d, \ldots )$                                       & \texttt{q.vector()}    & Associated $n-1$ vector                    & \\
		$2 \arctan(q/a)$                                           & \texttt{q.angle()}     & Rotation angle                             & $\dagger$ \\
		$(b,c,d) / \sin(\theta/2)$                                 & \texttt{q.axis()}      & Rotation axis                              & $\dagger$ \\
		$\mathbf{q} \mathbf{v} \mathbf{q}^{-1}$                    & \texttt{q.rotate(v)}   & Rotation of a vector                       & \\
		$\mathcal{M}$                                              & \texttt{q.matrix<n>()} & Matrix representation                      & \\
		$a = 0$                                                    & \texttt{q.isPure()}    & Real part is zero                          & \\
		$(b, c, d, \ldots ) = 0$                                   & \texttt{q.isScalar()}  & Unreal part is zero                        & \\
		\midrule
		$ \textbf{q} = \textbf{p}$								   & \texttt{q.isNear(p)}   & Are the same & \\
		$ \textbf{q} = 0$										   & \texttt{q.isNull()}    & Is zero & \\
																   & \texttt{q.isNan()}     & Has any NaN component  & \\
																   & \texttt{q.isNormal()}  & All components are neither infinity, NaN, zero or subnormal & \\
		$ \textbf{q} \ne \pm \infty$						       & \texttt{q.isFinite()}  & All components are finite & \\
		$ \textbf{q} = \pm \infty$								   & \texttt{q.isInfinite()}& Has any infinite component & \\
		\bottomrule
	\end{tabular}
	\caption{\noindent List of functions implemented for complex, quaternions, octonions, and sedenions, in addition to the one presented in \cref{tab:common_operations}. Methods marked with $\dagger$ are applicable to quaternions only. The $\ast$ marks the availability of an extra operation, which starts with \texttt{do} prefix (e.g. \texttt{doConj}), that works inplace on the quantity.}
	\label{tab:nions_methods}
\end{table}

\begin{table}
	\centering
	\begin{tabular}{p{.25\linewidth} p{.60\linewidth} r}
		\toprule
		Operator        & Operation                                                                  & Note \\
		\midrule
		\texttt{u + v}  & $(u_1+v_1, u_2+v_2, \ldots, u_n+v_n)$                                      & \\
		\texttt{u - v}  & $(u_1-v_1, u_2-v_2, \ldots, u_n-v_n)$                                      & \\
		\midrule
		\texttt{u * v}  & \textit{(Hamilton multiplication)}                                         & $\dagger$ \\
		\texttt{v * k}  & $(v_1\:k,v_2\:k, \ldots, v_n\:k)$                                          &\\
		\texttt{k * v}  & $(k\:v_1,k\:v_2, \ldots, k\:v_n)$                                          &\\
		\midrule
		\texttt{u / v}  & \textit{(Hamilton division)}                                               & $\dagger$ \\
		\texttt{v / k}  & $(v_1/k, v_2/k, \ldots, v_n/k )$                                           & \\
		\midrule
		\texttt{u >  v} & $u_1^2 + u_2^2 + \ldots + u_n^2 > v_1^2 + v_2^2 + \ldots + v_n^2 $         & \\
		\texttt{u <  v} & $u_1^2 + u_2^2 + \ldots + u_n^2 < v_1^2 + v_2^2 + \ldots + v_n^2 $         & \\
		\texttt{u >= v} & $u_1^2 + u_2^2 + \ldots + u_n^2 \ge v_1^2 + v_2^2 + \ldots + v_n^2 $       & \\
		\texttt{u <= v} & $u_1^2 + u_2^2 + \ldots + u_n^2 \le v_1^2 + v_2^2 + \ldots + v_n^2 $       & \\
		\midrule
		\texttt{u == v} & $u_1=v_1 \textrm{ and } u_2=v_2 \; \ldots \textrm{ and } u_n=v_n$          & \\
		\texttt{u != v} & $u_1 \ne v_1 \textrm{ or } u_2 \ne v_2 \;\ldots \textrm{ or } u_n \ne v_n$ & \\
		\bottomrule
	\end{tabular}
	\caption{\noindent List of operators implemented for each type of $n$-dimensional vector, quaternions, octonions, sedenions, and complex numbers. The $\dagger$ symbol marks hamiltonian multiplication and division, operations available only for quaternions, octonions, sedenions, and complex numbers. Relational operators compare the module of the vectors at the first instance: if the module is equal, then the vector's elements are compared one by one to obtain a strong ordering. }
	\label{tab:vector_operators}
\end{table}

\subsection{Frame transformations} \label{sec:ref_transformation}
While we were using \texttt{Univec} we found that some formulas become much simpler when we use specific coordinates transformation. One good example comes from the simulation of microscopic electron-molecule non-relativistic interactions in the center-of-momentum reference frame. A transformation-matrix can represent any linear transformation: complex transformations can be easily represented and computed by chaining multiple simpler transformations such as translations, rotations, and scaling operations.
\begin{table}
	\centering
	\begin{tabular}{p{.25\textwidth}p{.68\textwidth}}\toprule
		Class name & Description \\
		\midrule
		\texttt{TranslateFrame}  &  Linear translation by a compatible vector  \\
		\texttt{RotateFrame2D}     &  $2D$ Rotation (by angle or complex) \\
		\texttt{RotateFrame3D}     &  $3D$ Rotation (by Euler angles or quaternion) \\
		\texttt{ScaleFrame}      &  Scaling by a scalar or by a different value for each axis  \\
		\texttt{CompositeFrame}  &  Wrap two transformation in a single one  \\
		\bottomrule
	\end{tabular}
	\caption{\noindent List of ready to use reference frame transformations, that can be used to describe complex frame transformations. The class \texttt{CompositeFrame} is rarely used directly but can be created combining the other transformations using the \texttt{>>} operator.}
	\label{tab:classes-frame}
\end{table}

One can make a frame transformation using a concrete implementation of the abstract class \texttt{BaseFrameC3D} (in \cref{tab:classes-frame}, we present ready-to-use implementations).

A vector can be easily transformed into a new reference frame:
\begin{minted}{c++}

VectorC3D<QtySiLength>        translation(0. * meter, 1. * meter, 0. * meter);
TranslateFrame<3,QtySiLength> frame(translation);
VectorC3D<QtySiLength>        globalPosition;                                // ( 0, 0, 0) m
VectorC3D<QtySiLength>        localPosition = frame.forward(globalPosition); // ( 0,-1, 0) m
localPosition += VectorC3D<QtySiLength>(1. * meter, 1. * meter, 1. * meter); // ( 1, 0, 1) m
globalPosition = frame.backward(localPosition);                              // ( 1, 1, 1) m
\end{minted}
Multiple transformations can be combined to describe complex scenarios using the syntax:
\begin{minted}{c++}

VectorC3D<QtySiLength> v1; // (0,0,0) m
VectorC3D<QtySiLength> v2(0. * meter, -sqrt(2) * meter, sqrt(2) * meter);

VectorC3D<QtySiLength>       translation(0. * meter, 1. * meter, 0. * meter);
EulerRotation3D<QtySiLength> rotation(
        45. * degrees, 0. * degrees, 0. * degrees,
        RotationMode::INTRINSIC,
        RotationAxis::X, RotationAxis::Y, RotationAxis::Z);
QtySiDimensionless           scale(0.5);

auto frame = TranslateFrame<3,QtySiLength>(translation)
        >> RotateFrame3D<QtySiLength>(rotation)
        >> ScaleFrame<3,QtySiLength>(scale);

assert(frame.forward(v1).isNear(v2));
assert(frame.backward(v2).isNear(v1));
\end{minted}

\subsection{Vector conversion} \label{sec:coord_sys_transformation}
Another valuable feature of \texttt{Univec} is changing the coordinate system of a vector. The conversion from a type to another is made via a constructor, which is explicit when the conversion is computationally expensive (e.g. \texttt{VectorP2D} to \texttt{VectorC<2>}), and implicit otherwise (e.g. \texttt{VectorC<4>} to \texttt{Quaternion}).

These methods can be useful for executing operations on vectors represented in different coordinate systems.
For example, for converting a \texttt{VectorP2D} into a \texttt{VectorC2D}, we can use the following code:
\begin{minted}{c++}
VectorP2D<QtySiLength> vec_a(5. * meters,  90. * degrees);

// Cartesian coordinates
VectorC2D<QtySiLength> vec_b(vec_a); // ( 0, 5) m
\end{minted}
while for $3D$ vectors, we can use:
\begin{minted}{c++}
VectorC3D<QtySiLength> vec_a(3. * meters, 4. * meters, 12. * meters);

// Spherical coordinates
VectorS3D<QtySiLength> vec_b(vec_a); // ( 13 m, 22 deg, 53 deg)

// Cylindrical coordinates
VectorY3D<QtySiLength> vec_c(vec_a); // ( 5 m, 53 deg, 12 m)
\end{minted}
and for complex and quaternions, we have:
\begin{minted}{c++}
Complex<QtySiFrequency> c(1. * hertz,  2. * hertz);
Quaternion<QtySiLength> q(1. * meters, 2. * meters, 3. * meters, 4. * meters);

// 2x2 matrix
Matrix<2,2,QtySiFrequency> m1 = c.matrix();

// 4x4 matrix
Matrix<4,4,QtySiLength>    m2 = q.matrix();
\end{minted}

\begin{table}
	\centering
	\begin{tabular}{p{.25\linewidth} p{.40\linewidth} p{.25\linewidth}}\toprule
		Name                 & Description                              & Alias of \\
		\midrule
		\texttt{VectorC<N>}  & Generic $n$-dimensional Cartesian vector & \\
		\texttt{VectorC1D}   & Alias for $1D$ vectors                   & \texttt{VectorC<1>} \\
		\texttt{VectorC2D}   & Alias for $2D$ vectors                   & \texttt{VectorC<2>} \\
		\texttt{VectorC3D}   & Alias for $3D$ vectors                   & \texttt{VectorC<3>} \\
		\midrule
		\texttt{VectorP2D}   & Vector $2D$ in polar coordinates         & \\
		\texttt{VectorS3D}   & Vector $3D$ in spherical coordinates     & \\
		\texttt{VectorY3D}   & Vector $3D$ in cylindrical coordinates   & \\
		\midrule
		\texttt{Real}        & Real quantity                            & \texttt{VectorC<-1>}  \\
		\texttt{Complex}     & Complex quantity                         & \texttt{VectorC<-2>}  \\
		\texttt{Quaternion}  & Quaternion quantity                      & \texttt{VectorC<-4>}  \\
		\texttt{Octonion}    & Octonion quantity                        & \texttt{VectorC<-8>}  \\
		\texttt{Sedenion}    & Sedenion quantity                        & \texttt{VectorC<-16>} \\
		\midrule
		\texttt{Matrix<M,N>} & $M \times N$ generic matrix              & \\
		\midrule
		\texttt{RMatrix2}     & $2 \times 2$ square real-matrix               & \texttt{Matrix<2,2,Real>} \\
		\texttt{RMatrix3}     & $3 \times 3$ square real-matrix               & \texttt{Matrix<3,3,Real>} \\
		\texttt{RMatrix4}     & $4 \times 4$ square real-matrix               & \texttt{Matrix<4,4,Real>} \\
		\midrule
		\texttt{CMatrix2}     & $2 \times 2$ square complex-matrix            & \texttt{Matrix<2,2,Complex>}    \\
		\texttt{CMatrix3}     & $3 \times 3$ square complex-matrix            & \texttt{Matrix<3,3,Complex>} \\
		\texttt{CMatrix4}     & $4 \times 4$ square complex-matrix            & \texttt{Matrix<4,4,Complex>} \\
		\midrule
		\texttt{HMatrix2}     & $2 \times 2$ square quaternion-matrix         & \texttt{Matrix<2,2,Quaternion>} \\
		\texttt{HMatrix3}     & $3 \times 3$ square quaternion-matrix         & \texttt{Matrix<3,3,Quaternion>} \\
		\texttt{HMatrix4}     & $4 \times 4$ square quaternion-matrix         & \texttt{Matrix<4,4,Quaternion>} \\
		\midrule
		\texttt{OMatrix2}     & $2 \times 2$ square octonion-matrix         & \texttt{Matrix<2,2,Octonion>} \\
		\texttt{OMatrix3}     & $3 \times 3$ square octonion-matrix         & \texttt{Matrix<3,3,Octonion>} \\
		\texttt{OMatrix4}     & $4 \times 4$ square octonion-matrix         & \texttt{Matrix<4,4,Octonion>} \\
		\midrule
		\texttt{SMatrix2}     & $2 \times 2$ square sedenion-matrix         & \texttt{Matrix<2,2,Sedenion>} \\
		\texttt{SMatrix3}     & $3 \times 3$ square sedenion-matrix         & \texttt{Matrix<3,3,Sedenion>} \\
		\texttt{SMatrix4}     & $4 \times 4$ square sedenion-matrix         & \texttt{Matrix<4,4,Sedenion>} \\
		\bottomrule
	\end{tabular}
	\caption{List of classes and aliases implemented in \texttt{Univec}.}
	\label{tab:classes_and_aliases}
\end{table}

\begin{table*}
	\centering
	\begin{tabular}{p{.18\textwidth} p{.30\textwidth} p{.33\textwidth} r}\toprule
		Operation                                                 & Method                            & Description                                                        & Note \\
		\midrule
		$\mathcal{A}_{i-1,j-1}$                                   & \texttt{a(i,j)}                   & Element access ($1$-based)                                         & \\
		$\mathcal{A}_{i,j}$                                       & \texttt{a[i,j]}                   & Element access ($0$-based)                                         & \\
									        					  & \texttt{a.rowMajor(x)}            & Element access (row-major)                                         & \\
		                                					      & \texttt{a.colMajor(x)}            & Element access (column-major)                                      & \\
		\midrule
		$m$                                                       & \texttt{a.rows}                   & Rows count                                                         & \\
		$n$                                                       & \texttt{a.columns}                & Columns count                                                      & \\
		$\mathcal{A}_{i,*}$                                                          & \texttt{a.row()}                  & Returns a specified row                         & \\
		$\mathcal{A}_{*,j}$                                                         & \texttt{a.col()}                  & Returns a specified column                       & \\
		$m=n$                                                     & \texttt{a.square}                 & Check if square                                                    & \\
		\midrule
		$ \mathcal{A}_{i_2,*} \leftarrow \mathcal{A}_{i_2,*} + \mathcal{A}_{i_1,*}$             & \texttt{a.rowAdd(i1, i2)}			& Row addition and store               & $\ast$ \\
		$ \mathcal{A}_{i_2,*} \leftarrow \mathcal{A}_{i_2,*} - \mathcal{A}_{i_1,*}$             & \texttt{a.rowSub(i1, i2)}         & Row subtraction and store            & $\ast$ \\
		$ \mathcal{A}_{i,*} \leftarrow \mathcal{A}_{i,*} \cdot s$ 								& \texttt{a.rowMul(i, s)}        	& Row multiplication                   & $\ast$ \\
		$ \mathcal{A}_{i,*} \leftarrow \mathcal{A}_{i,*} / s$ 							     	& \texttt{a.rowDiv(i, s)} 			& Row division                         & $\ast$ \\
		$ \mathcal{A}_{i_1,*} \leftrightarrow \mathcal{A}_{i_2,*}$ 				    			& \texttt{a.rowSwap(i1, i2)}        & Row swap                             & $\ast$\\
		                                                          & \texttt{a.reverseRows()}          & Reverse the rows                                                   & $\ast$\\
		                                                          & \texttt{a.shuffleRows()}          & Shuffles the rows                                                  & $\ast$\\
		                                                          & \texttt{a.sortRows()}             & Sorts the rows                                                     & $\ast$\\
		\midrule                                         
		$ \mathcal{A}_{*,j_2} \leftarrow \mathcal{A}_{*,j_2} + \mathcal{A}_{*,j_1}$                   & \texttt{a.colAdd(j1, j2)} & Column addition and store              & $\ast$ \\
		$ \mathcal{A}_{*,j_2} \leftarrow \mathcal{A}_{*,j_2} - \mathcal{A}_{*,j_1}$                   & \texttt{a.colSub(j1, j2)} & Column subtraction and store           & $\ast$ \\
		$ \mathcal{A}_{*,j} \leftarrow \mathcal{A}_{*,j} \cdot s$                                     & \texttt{a.colMul(j, s)} & Column multiplication                    & $\ast$ \\
		$ \mathcal{A}_{*,j} \leftarrow \mathcal{A}_{*,j} / s$                                         & \texttt{a.colDiv(j, s)} & Column division                          & $\ast$ \\
		$ \mathcal{A}_{*,j_1} \leftrightarrow \mathcal{A}_{*,j_2}$          		                  & \texttt{a.colSwap(j1, j2)}   & Column swap                         & $\ast$\\
		                                                          & \texttt{a.reverseCols()}          & Reverse the columns                                				   & $\ast$\\
		                                                          & \texttt{a.shuffleCols()}          & Shuffle the columns                                                & $\ast$\\
		                                                          & \texttt{a.sortCols()}             & Sorts the columns                                                  & $\ast$\\
        \midrule
		$o(\mathcal{A})$                                          & \texttt{a.order()}                & Order                                                              & $\dagger$ \\
		$O(\mathcal{A})$                                          & \texttt{a.orders()}               & Order (in $m \times n $ form)                                      & \\
		$\text{adj}(\mathcal{A})$                                 & \texttt{a.adj()}                  & Ajugate                                                            & $\dagger$ \\
		$c_{m,n}$                                                 & \texttt{a.cofactor(m,n)}          & Cofactor element                                                   & $\dagger$ \\
		$C_{m,n}$                                                 & \texttt{a.cofactors(m,n)}         & Cofactor matrix                                                    & $\dagger$ \\
		$\det(\mathcal{A})$                                       & \texttt{a.det()}                  & Determinant                                                        & $\dagger$ \\
		$\text{tr}(\mathcal{A})$                                  & \texttt{a.tr()}                   & Trace                                                              & $\dagger$ \\
		$\mathcal{A}^T$                                           & \texttt{a.t()}                    & Transpose                                                          & $\ast$\\
		$\overline{A}$                                            & \texttt{a.conj()}                 & Conjugate                                                          & $\ast$\\
		$\mathcal{A}^H$                                           & \texttt{a.conjT()}                & Conjugate transpose                                                & $\ast$\\
		$\mathcal{A}_{i,j} \leftarrow 0 \quad \text{if} |{A}_{i,j}| < t $ & \texttt{a.zap(t)} 		  & Set to zero small elements								           & $\ast$\\
		$\mathcal{A}^{-1}$                                        & \texttt{a.inv()}                  & Inverse                                                            & $\dagger$ \\
		$\mathcal{A}^{+}$                                         & \texttt{a.pseudoInv()}            & Moore-Penrose inverse                                              & \\
		$\mathcal{A}[m;n]$                                        & \texttt{a.sub(n,m)}               & Submatrix                                                          & \\
		$m_{m,n}$                                                 & \texttt{a.minor(m,n)}             & Minor element                                                      & $\dagger$ \\
		$M_{m,n}$                                                 & \texttt{a.minors(m,n)}            & Minor matrix                                                       & $\dagger$ \\
		$(\mathcal{A}_{1,1},\mathcal{A}_{2,2}, \ldots, \mathcal{A}_{n,n})$   			    & \texttt{a.diagonal()}             & Diagonal elements                        & $\dagger$ \\
		$\mathcal{A}_{1,1} \cdot \mathcal{A}_{2,2} \cdot \ldots \cdot \mathcal{A}_{n,n})$   & \texttt{a.diagonalProduct()}      & Product of the diagonal elements         & $\dagger$ \\
		$\langle \mathcal{A}, \mathcal{B} \rangle_F$              & \texttt{a.frobenius(b)}           & Frobenius inner product                                            & \\
		$\sum \mathcal{A}_{i,j}$                                  & \texttt{a.grandSum()}             & Sum of all elements                    						       & \\
		$\mathcal{A} \oplus  \mathcal{B}$                         & \texttt{a.directSum(b)}           & Direct sum                              						   & \\
		$\mathcal{A} \otimes \mathcal{I}_b + \mathcal{I}_a \otimes \mathcal{B}$                       & \texttt{a.kroneckerSum(b)}        & Kronecker sum                                                      & \\
		$\mathcal{A} \otimes \mathcal{B}$                         & \texttt{a.kroneckerProduct(b)}    & Kronecker product                                                  & \\
		$\mathcal{A}_{i,j} \cdot \mathcal{B}_{i,j}$               & \texttt{a.elementwiseProduct()}   & Element-wise product                                             & \\
		$\mathcal{A}_{i,j} / \mathcal{B}_{i,j}$                   & \texttt{a.elementwiseDivision()}  & Element-wise divisison                                             & \\
		                                                          & \texttt{a.reduction(type)}        & Performs the reduction of the matrix                               & $\ast$\\

		\bottomrule
	\end{tabular}
	\caption{List of functions implemented for the \texttt{Matrix<M,N>} class. The $\dagger$ symbol marks operations available for square matrix only, while $\ddagger$ symbols is available for row or column matrix only. The $\ast$ marks the disponibility of an extra operation, which starts with \texttt{do} prefix (e.g. \texttt{doColSwap}), that works inplace on the matrix. Matrix indices are zero-based.}
	\label{tab:matrix_methods}
\end{table*}

\begin{table*}
	\centering
	\begin{tabular}{p{.18\textwidth} p{.30\textwidth} p{.33\textwidth} r}\toprule
		Operation                                                 & Method                          & Description                                & \\
		\midrule
		$ \mathcal{A}_{m,n} < \infty$                             & \texttt{a.isFinite()}           & All elements are finite                    & \\
		$ \mathcal{A}_{m,n} = \pm\infty$                          & \texttt{a.isInfinite()}         & Any element is not finite                  & \\
		$ \mathcal{A}_{m,n} \notin \mathbb{R} $                   & \texttt{a.isNan()}              & Any element is a \texttt{nan}              & \\
		$ \mathcal{A} \approx \mathcal{B} $                       & \texttt{a.isNear(b)}            & Is approximate equals                      & \\
		$ \mathcal{A}_{m,n} \in \mathbb{R} \textbackslash \{0\} $ & \texttt{a.isNormal()}           & All elements are finite, real and non-zero & \\
		$ \mathcal{A}_{m,n} = 0$                                  & \texttt{a.isNull()}             & All elements are zero                      & \\
		\midrule
		$\mathcal{A} = \mathcal{I}$                               & \texttt{a.isIdentity()}         & Is identity                                & $\dagger$ \\
		$\mathcal{A}^H = \mathcal{A}^{-1}$                        & \texttt{a.isUnitary()}          & Is unitary                                 & $\dagger$ \\
		$\mathcal{A}_{m,n} =  \overline{\mathcal{A}}_{n,m}$       & \texttt{a.isHermitian()}        & Is hermitian                               & $\dagger$ \\
		$\mathcal{A}_{m \neq n} = 0$                              & \texttt{a.isDiagonal()}         & Is diagonal                                & $\dagger$ \\
		$\mathcal{A} A^T = \mathcal{I}$                           & \texttt{a.isOrthogonal()}       & Is orthogonal                              & $\dagger$ \\
		$\det(\mathcal{A}) = 0$                                   & \texttt{a.isSingular()}         & Is singular                                & $\dagger$ \\
		$\mathcal{A}_{m,n} = \mathcal{A}_{n,m}$                   & \texttt{a.isSymmetric()}        & Is symmetric                               & $\dagger$ \\
		$\mathcal{A}_{m,n} = -\overline{\mathcal{A}}_{n,m}$       & \texttt{a.isSkewHermitian()}    & Is skew-hermitian                          & $\dagger$ \\
		$\mathcal{A}_{m,n} = -\mathcal{A}_{n,m}$                  & \texttt{a.isSkewSymmetric()}    & Is skew-symmetric                          & $\dagger$ \\
		$\mathcal{A}_{m,n} = -\mathcal{A}_{n,m}$                  & \texttt{a.isTriangular()}       & Is triangular                              & $\dagger$ \\
		$\mathcal{A}_{m < n} = 0$                                 & \texttt{a.isLowerTriangular()}  & Is lower-triangular                        & $\dagger$ \\
		$\mathcal{A}_{m > n} = 0$                                 & \texttt{a.isUpperTriangular()}  & Is upper-triangular                        & $\dagger$ \\
		                                                          & \texttt{a.isStrictTriangular()} & Is strict-triangular                       & $\dagger$ \\
		                                                          & \texttt{a.isRowEchelon()}       & Is row echelon form                        & \\
		                                                          & \texttt{a.isColEchelon()}       & Is column echelon form                     & \\
		                                                          & \texttt{a.isRedRowEchelon()}    & Is reduced row echelon form                & \\
		                                                          & \texttt{a.isRedColEchelon()}    & Is reduced column echelon form             & \\
		\midrule
		$\|\mathcal{A}\|$                                         & \texttt{a.norm()}                 &Euclidean row or column norm                                        & $\ddagger$ \\
		$\|\mathcal{A}\|^2$                                       & \texttt{a.normSquared()}          &Euclidean row or column norm squared                                & $\ddagger$ \\
		$\|\mathcal{A}\|_{1}$                                     & \texttt{a.normL1()}               & $L_1$ norm                                                         & \\
		$\|\mathcal{A}\|_{max}$                                   & \texttt{a.normLMax()}             & Max norm                                                           & \\
		$\|\mathcal{A}\|_{\infty}$                                & \texttt{a.normLInf()}             & Infinity norm                                                      & \\
		$\|\mathcal{A}\|_{p,q}$                                   & \texttt{a.normL(p,q)}             & $pq$ norm                                                          & \\
		$\|\mathcal{A}\|_{F}$                                     & \texttt{a.normF()}                & Frobenius norm                                                     & \\
		\bottomrule
	\end{tabular}
	\caption{List of functions implemented for the \texttt{Matrix<M,N>} class. The $\dagger$ symbol marks operations available for square matrix only, while $\ddagger$ symbols is available for row or column matrix only.}
	\label{tab:matrix_is_methods}
\end{table*}

\begin{table*}
	\centering
	\begin{tabular}{p{.16\textwidth} p{.16\textwidth} p{.55\textwidth}}\toprule
		Operation & Operator & Description \\
		\midrule
		$\mathcal{A} + \mathcal{B}$ & \texttt{a +  b} & Sum    \\
		$\mathcal{A} - \mathcal{B}$ & \texttt{a -  b} & Subtraction    \\

		$\mathcal{A} \; \mathcal{B}$ & \texttt{a * b} & Multiplication   \\
		$\mathcal{A} \; c$ & \texttt{a * c} & Multiplication with a scalar    \\
		$c \; \mathcal{A}$ & \texttt{c * a} & Multiplication with a scalar    \\
		$\mathcal{A} \; \textbf{v}$ & \texttt{a * v} & Multiplication with a vector   \\
		$\mathcal{A} / c$ & \texttt{a / c} & Division by a scalar    \\
		\bottomrule
	\end{tabular}
	\caption{List of functions implemented for the \texttt{Matrix<M,N>} class.}
	\label{tab:matrix_operations}
\end{table*}

\subsection{Usage examples} \label{sec:usage}

In this subsection, we will present two realistic examples of using \texttt{Univec}. For the first example we will show a simple implementation of the Bethe-Bloch equation, a handy formula for evaluating the energy deposit of ionizing charged particles. For the second example we will show how our library can be used in elementary particle physics, evaluating the decay rate of the $W^-$ boson. It is important to notice that we perform the calculation using SI units and not the natural units that are normally used. This is because there is currently no support for natural units in \texttt{Boost::Units}, but we plan to address this issue in the near future (see \cref{sec:limits}). All the examples presented in this article are available in a public repository\cite{univec_examples_web}.

\subsubsection{Bethe-Bloch}

The well-known Bethe-Bloch formula\cite{bethe_zur_1930,bloch_zur_1933,bloch_bremsvermogen_1933} is used to calculate the mean energy loss of a charged particle. For example, if we have a particle with velocity $v$, charge $z$ (in a unit of elementary charges), traveling a medium with electron number density $n$ and mean excitation energy $I$, we can calculate the mean energy-loss using the formula:
\begin{align}
-\left\langle\frac{d E}{d x}\right\rangle=\frac{4 \pi}{m_e c^2} \cdot \frac{n z^2}{\beta^2} \cdot\left(\frac{e^2}{4 \pi \varepsilon_0}\right)^2 \cdot\left[\ln \left(\frac{2 m_e c^2 \beta^2 \gamma^2}{I }\right)-\beta^2\right]
\end{align}
where $m_e$ is the electron rest mass, $c$ is the speed of light and $\beta=v/c$ as well as $\gamma=1 / \sqrt{1 - \beta^2}$ are the Lorentz factors.

Using \texttt{Univec}, we can encode this formula into a compact expression, enjoying a fully dimensional analysis validation:
\begin{minted}{c++}

QtySiDimensionless molZ            = 10; // Neon
QtySiDimensionless eleChargeNumber = -1;
QtySiDimensionless pi (M_PI);
QtySiEnergy        eleRestEnergy (m_e * c * c);
QtySiEnergy        eleKinEnergy  (1000. * kilo * volt * e);
QtySiDensity       molDensity    (2.6867811e25 / cubic_meter);
QtySiDensity       eleDensity    (molZ * molDensity);
QtySiEnergy        meanExcEnergy (137. * volt * e);

QtySiDimensionless gamma  = (eleRestEnergy + eleKinEnergy) / eleRestEnergy;
QtySiDimensionless gamma2 = gamma * gamma;
QtySiDimensionless beta2  = 1. - 1. / gamma2;

auto               p1 = 4. * pi / (m_e * c * c);
QtySiDensity       p2 = eleDensity * eleChargeNumber * eleChargeNumber / beta2;
auto               p3 = e * e / (4. * pi * epsilon_0);
QtySiDimensionless p4 = log(2. * m_e * c * c * beta2 * gamma2 / meanExcEnergy) - beta2;

cout << "Mean energy loss: " 
     << (double) ((-p1 * p2 * p3 * p3 * p4) / (1e3 * volt * e / meter)) << " keV/m" << endl;
\end{minted}

\subsubsection{$W^- \rightarrow e^- + \bar{\nu}_e$ decay} \label{subsec:ex-particle}
The previous example was helpful for presenting the dimensional analysis validation, but the equation itself only contained scalar operation. We will show here how vector, spinor, and matrix operations can be combined in a compact syntax to perform advanced calculations. In this example we reproduce a derivation from \cite[ch.~15]{thomson_modern_2013} which describes the properties of the $W$ and $Z$ bosons. To calculate the decay rate of a $W^-$ boson into an electron, $e^-$, and an electronic anti-neutrino, $\overline{\nu}_e$ (shown in \cref{fig:fym-diag-1}),
\begin{figure}[h]
	\centering
	\begin{fmffile}{feynman-diag-1}
		\begin{fmfgraph*}(80,40)
			\fmfleft{p1}
			\fmfright{p3,p4}
			\fmf{boson,label=$p_1$}{p1,v}
			\fmf{fermion,label=$p_3$}{v,p3}
			\fmf{fermion,label=$p_4$}{p4,v}
			\fmfdot{v}
			\fmflabel{$W^-$}{p1}
			\fmflabel{$\overline{\nu}_e$}{p4}
			\fmflabel{$e^-$}{p3}
		\end{fmfgraph*}
	\end{fmffile}
	\caption{Lowest-order Feynman diagram for $W^-$ boson decay into an electron and an anti-neutrino\cite{thomson_modern_2013}.}
	\label{fig:fym-diag-1}
\end{figure}

we calculate the decay matrix elements using
\begin{align}\label{eq:matrix-elements}
	\mathcal{M}_{f i}=\frac{g_W}{\sqrt{2}} \epsilon_\mu^\lambda\left(p_1\right) \bar{u}\left(p_3\right) \gamma^\mu \frac{1}{2}\left(1-\gamma^5\right) v\left(p_4\right)
\end{align}
where $\gamma^\mu$ are the $\gamma$-matrices in Dirac-Pauli representation and $\epsilon_{\mu}^\lambda$ represents the three possible polarization states
\begin{align}
\epsilon_{-}^\mu=\frac{1}{\sqrt{2}}(0,1,-i, 0), \quad \quad \epsilon_L^\mu=\frac{1}{m_W}\left(p_z, 0,0, E\right), \quad \quad \epsilon_{+}^\mu=-\frac{1}{\sqrt{2}}(0,1, i, 0)
\end{align}
and  $v\left(p_4\right)$, $\bar{u}(p_3)$ are, respectively, the adjoint particle spinor and antiparticle spinor, defined as
\begin{align}
u_1(p)=N\left(\begin{array}{c}
	1 \\
	0 \\
	\frac{p_z}{E+m} \\
	\frac{p_x+i p_0}{E+m}
\end{array}\right), \quad u_2(p)=N\left(\begin{array}{c}
	0 \\
	1 \\
	\frac{p_x-i p_0}{E+m} \\
	\frac{-p_z}{E+m}
\end{array}\right), \quad v_1(p)=N\left(\begin{array}{c}
	\frac{p_x-i p_0}{E+m} \\
	\frac{-p_z}{E+m} \\
	0 \\
	1
\end{array}\right), \quad v_2(p)=N\left(\begin{array}{c}
	\frac{p_z}{E+m} \\
	\frac{p_x+i p_0}{E+m} \\
	1 \\
	0
\end{array}\right)
\end{align}
with $N=\sqrt{E + m}$. Finally, having $\mathcal{M}_{f i}$, we can calculate the average decay rate using
\begin{align}
\left\langle\left|\mathcal{M}_{f i}\right|^2\right\rangle=\frac{1}{3}\left(\left|\mathcal{M}_{-}\right|^2+\left|\mathcal{M}_L\right|^2+\left|\mathcal{M}_{+}\right|^2\right)
\end{align}
and
\begin{align}
	\Gamma = \frac{\mathrm{p}^*}{8 \pi m_{\mathrm{W}}^2}\left\langle\left|\mathcal{M}_{f i}\right|^2\right\rangle
\end{align}
where $p*$ is the momentum of the final particle in the center-of-mass frame.

The equations above are dimensional-invalid in SI units (e.g., we sum a mass and energy). However, in natural units, we define quantities like mass, momentum, length, etc., in equivalent energy units, so their operations are legit. Unfortunately, our current implementation does not support natural units, yet, so we are forced to convert to SI units to perform this calculation. The result is the code below, which uses our custom classes for matrices, vectors, and complex values to produce a compact and readable code with a full dimensional check at compile time:
\begin{minted}{c++}

QtySiEnergy eV = 1. * volt * e;
QtySiPlaneAngle pi = M_PI * radians;
auto h_bar = h / (2. * pi); // Energy * Time
QtySiDimensionless gW = e / (5.28e-19 * coulomb) / sqrt(0.231214);

QtySiEnergy P1_TotalEnergy (200.    * giga * eV);
QtySiEnergy P1_RestEnergy  ( 80.433 * giga * eV);
VectorC3D<> P1_Direction = VectorC3D<>::kZ();
// Using: E² = (pc)² + (m₀c²)²
VectorC3D<QtySiMomentum> P1_Momentum = sqrt(P1_TotalEnergy * P1_TotalEnergy - P1_RestEnergy * P1_RestEnergy) / c * P1_Direction;

QtySiPlaneAngle P3_Theta (45. * degrees);
VectorC3D<> P3_Direction = VectorC3D<>::kZ().rotateY(P3_Theta/2.);
VectorC3D<QtySiMomentum> P3_Momentum = P1_Momentum.norm() / 2. * P3_Direction;
QtySiEnergy P3_RestEnergy = m_e * c * c;
QtySiEnergy P3_TotalEnergy = sqrt(P3_RestEnergy * P3_RestEnergy + P3_Momentum.normSquared() * c * c);

QtySiPlaneAngle P4_Theta (-45. * degree::degrees);
VectorC3D<> P4_Direction = VectorC3D<>::kZ().rotateY(P4_Theta/2.);
VectorC3D<QtySiMomentum> P4_Momentum = P1_Momentum.norm() / 2. * P4_Direction;
QtySiEnergy P4_RestEnergy  (0.5 * eV);
QtySiEnergy P4_TotalEnergy = sqrt(P4_RestEnergy * P4_RestEnergy + P4_Momentum.normSquared() * c * c);

// P3 u1 spinor
auto P3_u1= sqrt(P3_TotalEnergy) * CMatrix<4,1> (
        Complex<>(1),
        Complex<>(0),
        Complex<QtySiEnergy>( P3_Momentum.z * c) / P3_TotalEnergy,
        Complex<QtySiEnergy>( P3_Momentum.x * c, P3_Momentum.y * c) / P3_TotalEnergy);

// P3 u2 spinor
auto P3_u2 = sqrt(P3_TotalEnergy) * CMatrix<4,1> (
        Complex<>(1),
        Complex<>(0),
        Complex<QtySiEnergy>( P3_Momentum.x * c, -P3_Momentum.y * c) / P3_TotalEnergy,
        Complex<QtySiEnergy>(-P3_Momentum.z * c) / P3_TotalEnergy);

// P4 v1 spinor
auto P4_v1 = sqrt(P4_TotalEnergy) * CMatrix<4,1> (
        Complex<QtySiEnergy>( P4_Momentum.x * c, -P4_Momentum.y * c) / P4_TotalEnergy,
        Complex<QtySiEnergy>(-P4_Momentum.z * c) / P4_TotalEnergy,
        Complex<>(0.),
        Complex<>(1.));

// P4 v2 spinor
auto P4_v2 = sqrt(P4_TotalEnergy) * CMatrix<4,1> (
        Complex<QtySiEnergy>( P4_Momentum.z * c) / P4_TotalEnergy,
        Complex<QtySiEnergy>( P4_Momentum.x * c, P4_Momentum.y * c) / P4_TotalEnergy,
        Complex<>(1.),
        Complex<>(0.));

auto fn_adjoint = [](const auto& spinor) { return spinor.conjT() * diracGamma<0>; };
auto fn_controvariant = [](const auto& tensor) { return CMetricTensorG<> * tensor; };

CMatrix<4,1> polMinus =  1./sqrt(2) * CMatrix<4,1>( Complex<>(), Complex<>(1), Complex<>(0,-1), Complex<>());
CMatrix<4,1> polPlus  = -1./sqrt(2) * CMatrix<4,1>( Complex<>(), Complex<>(1), Complex<>(0, 1), Complex<>());
CMatrix<4,1> polLong  = -1./P1_RestEnergy * CMatrix<4,1,QtySiEnergy>(
                                Complex<QtySiEnergy>(P1_Momentum.z * c),
                                Complex<QtySiEnergy>(),
                                Complex<QtySiEnergy>(),
                                Complex<QtySiEnergy>(QtySiEnergy(P1_TotalEnergy)));

std::map<int,CMatrix<4,1>> polarisations;
polarisations[-1] = polMinus;
polarisations[ 0] = polLong;
polarisations[+1] = polPlus;

typename power_typeof_helper<QtySiEnergy,static_rational<2>>::type  M1_avg, M2_avg;
for (int mu = 0; mu <= 3; mu++)
{
    for (int pol = -1; pol <= 1; pol++)
    {
        CMatrix<4,1,QtySiEnergy> M1 = gW / sqrt(2) * fn_controvariant(polarisations[pol]) *
                                      fn_adjoint(P3_u1) * diracGammaMu(mu) *
                                      0.5 * (CI<4> - diracGamma<5>) * P4_v1;

        CMatrix<4,1,QtySiEnergy> M2 = gW / sqrt(2) * fn_controvariant(polarisations[pol]) *
                                      fn_adjoint(P3_u2) * diracGammaMu(mu) *
                                      0.5 * (CI<4> - diracGamma<5>) * P4_v2;

        M1_avg += M1.normSquared() / 3.;
        M2_avg += M2.normSquared() / 3.;
    }
}

QtySiFrequency gamma1 =
        P3_Momentum.norm() * c / (8. * pi * P1_RestEnergy * P1_RestEnergy * h_bar) * M1_avg;
QtySiFrequency gamma2 =
        P3_Momentum.norm() * c / (8. * pi * P1_RestEnergy * P1_RestEnergy * h_bar) * M2_avg;

cout << "  Γ₁: " << gamma1 << "  τ₁: " << 1./gamma1 << endl;
cout << "  Γ₂: " << gamma2 << "  τ₂: " << 1./gamma2 << endl;
\end{minted}

As we have already explained, using SI units is not a natural choice for HEP physicists. Due to the nature of the subject, we have to handle SI units values that differ from each other by many orders of magnitude. Due to this fact, a substantial precision loss can occur during calculations (see \cref{sec:limits} for details). However, our example presented above produced pretty good results, near to the accepted value of $\tau \approx O(10^{-25}s)$.

\section{Methods}\label{sec:methods}
An effective software library needs an efficient and reliable procedure for validating both the functionality, the correctness, and the performances obtained. Therefore, in implementing this library, we perform two type of tests: a test of the functionality and correctness of the implemented methods and a test of performance compared with similar techniques for calculating the same operation.

When we talk about the functionality and performance tests, even though the source code of this library was carefully written, we felt the need to add an additional layer of verification by integrating a testing system.

To deliver a reliable application, we have defined a good software development practice by incorporating a Test-Driven Development (TDD) system. More details about this integration are presented in \cref{sec:testing}. Additionally, we chose to use two different compilers, \texttt{GCC}\cite{gcc} and \texttt{Clang}\cite{clang}, to support the wide range of machines our software could run on. The amount of code executed during our tests is greater than \SI{95}{\percent} (according to \texttt{gcovr}\cite{gcovr} - the code coverage utility we use), ensuring good code coverage.

\subsection{Unit testing}\label{sec:testing}

\texttt{Univec} was developed using the Test-Driven Development (TDD) software design paradigm where code development and testing occur simultaneously. By integrating a unit testing strategy, we can separately test for the correctness of our application's functionality. To implement these tests, \texttt{Univec} uses \texttt{GoogleTest}\cite{googletest}, one of the most popular \texttt{C++} unit testing frameworks. For each implemented method and operator of our code an associated test ensures that the expected behavior satisfies our specifications and that functionality is preserved.

We have chosen to integrate CI/CD, another modern software development practice, through automated integration of testing and code documentation into the stages of the software development. We use \texttt{GitLab}\cite{gitlab} to host our repository, allowing us to access the provided CI/CD feature. For each commit or push GitLab triggers the build of code, runs all unit tests implemented with \texttt{GoogleTest} and extracts the annotated \texttt{C++} files to generate the online \texttt{Univec} documentation using \texttt{Doxygen}\cite{doxygen}. Furthermore, at the end of each test procedure on the pipeline, several report files are generated: two files with the test results (one for each compiler: \texttt{junit\_gcc.xml} and \texttt{junit\_clang.xml}) and one file that briefly presents code coverage results for the tests run (\texttt{coverage\_gcc.xml}).

This combined approach, CI/CD and unit tests, assures us of the \texttt{Univec}'s functionality and maintainability.

\subsection{Performance study} \label{sec:performance}

\begin{figure*}
	\centering
	\subfigure[$2D$ vectors on Linux]{\includegraphics[width=0.49\textwidth]{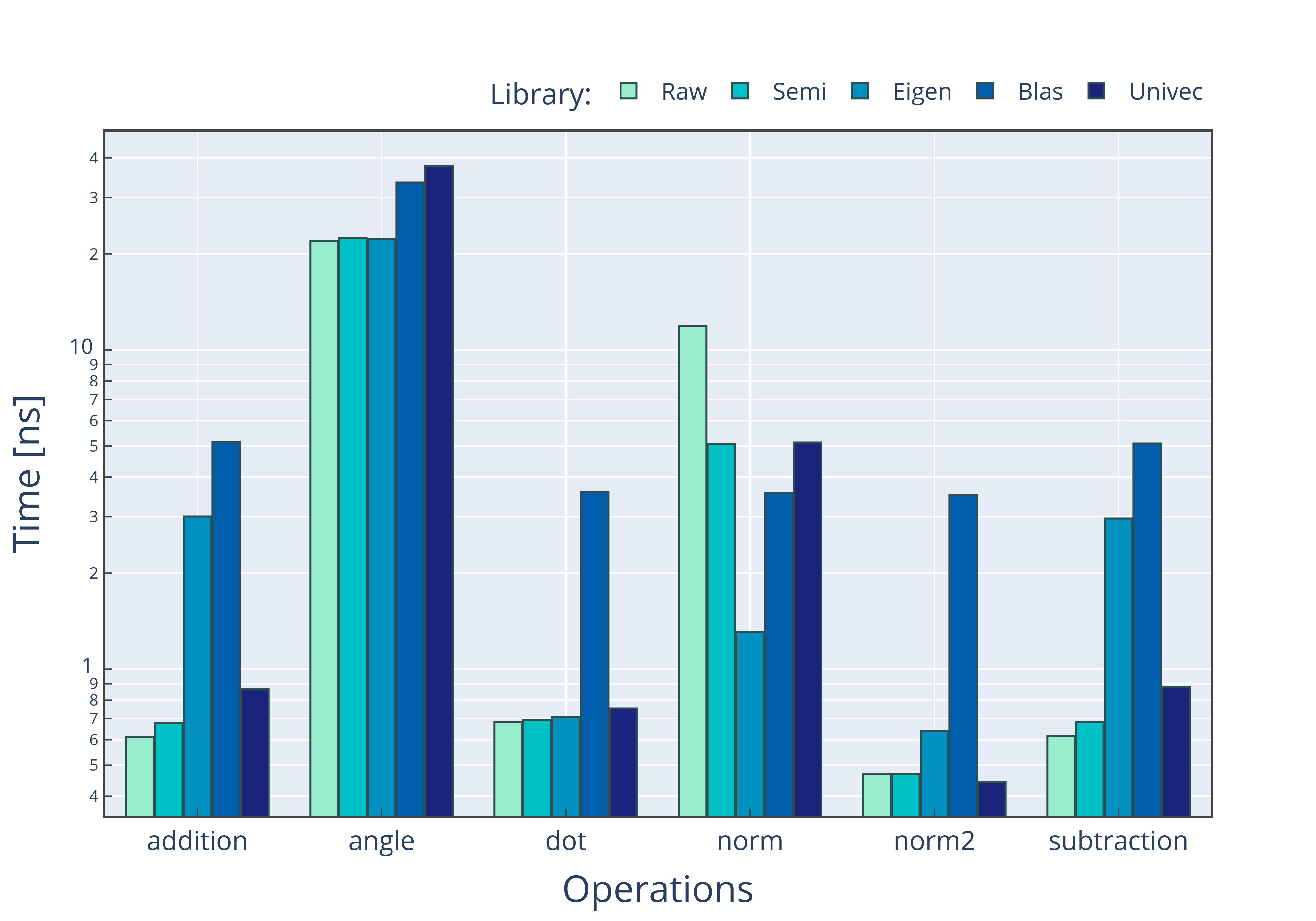}}
	\subfigure[$2D$ vectors on macOS]{\includegraphics[width=0.49\textwidth]{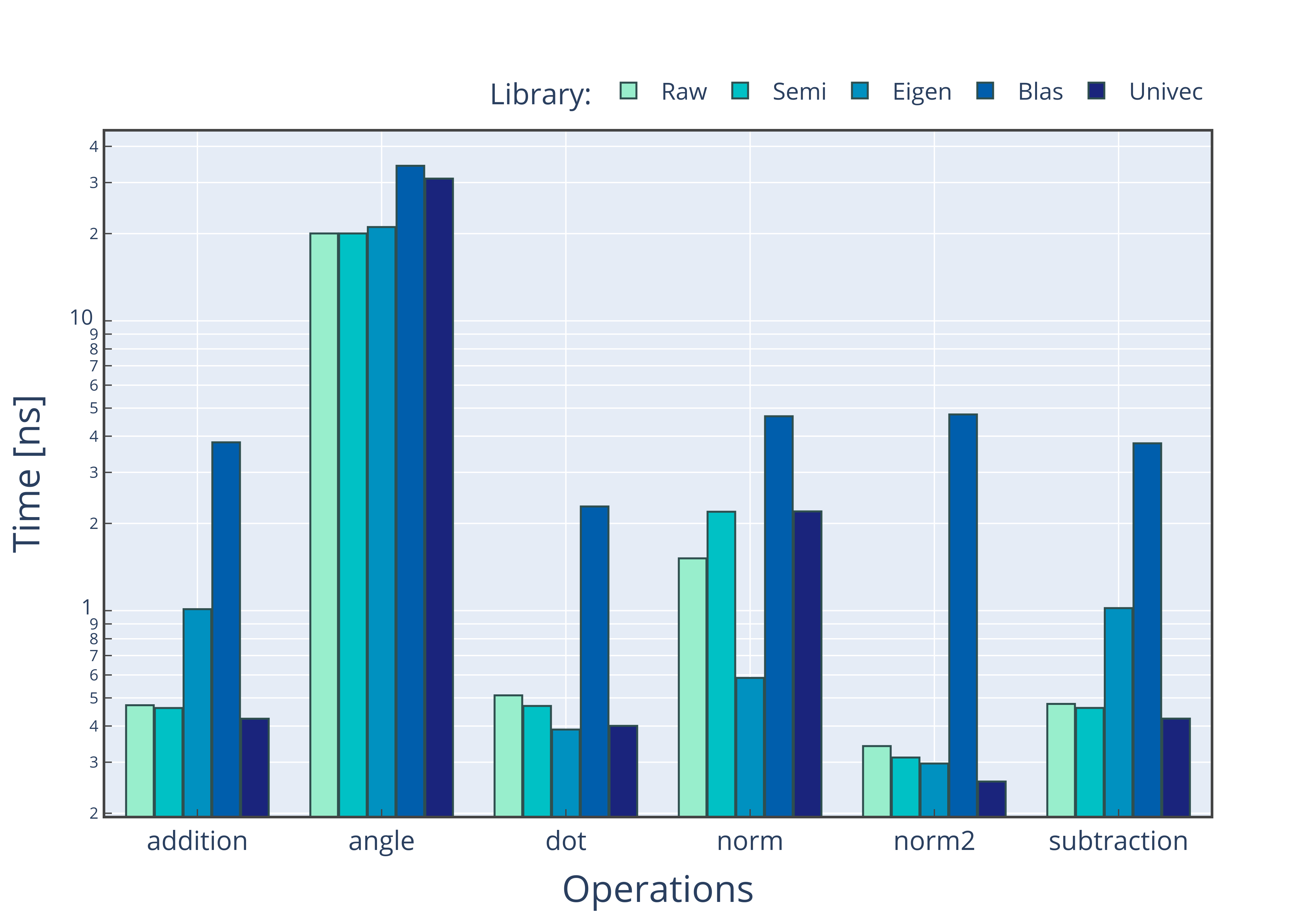}}
	
	\subfigure[$3D$ vectors on Linux]{\includegraphics[width=0.49\textwidth]{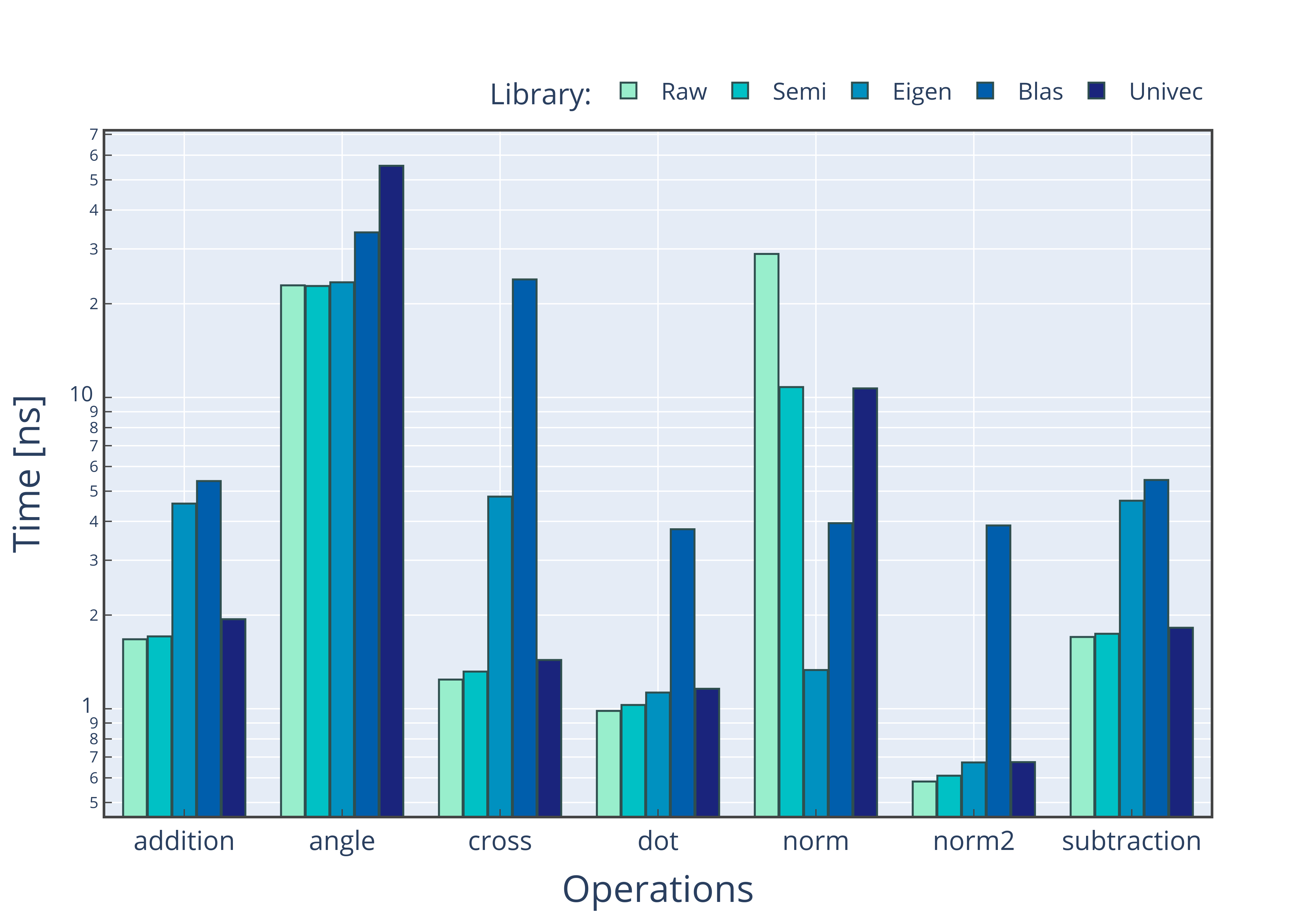}}
	\subfigure[$3D$ vectors on macOS]{\includegraphics[width=0.49\textwidth]{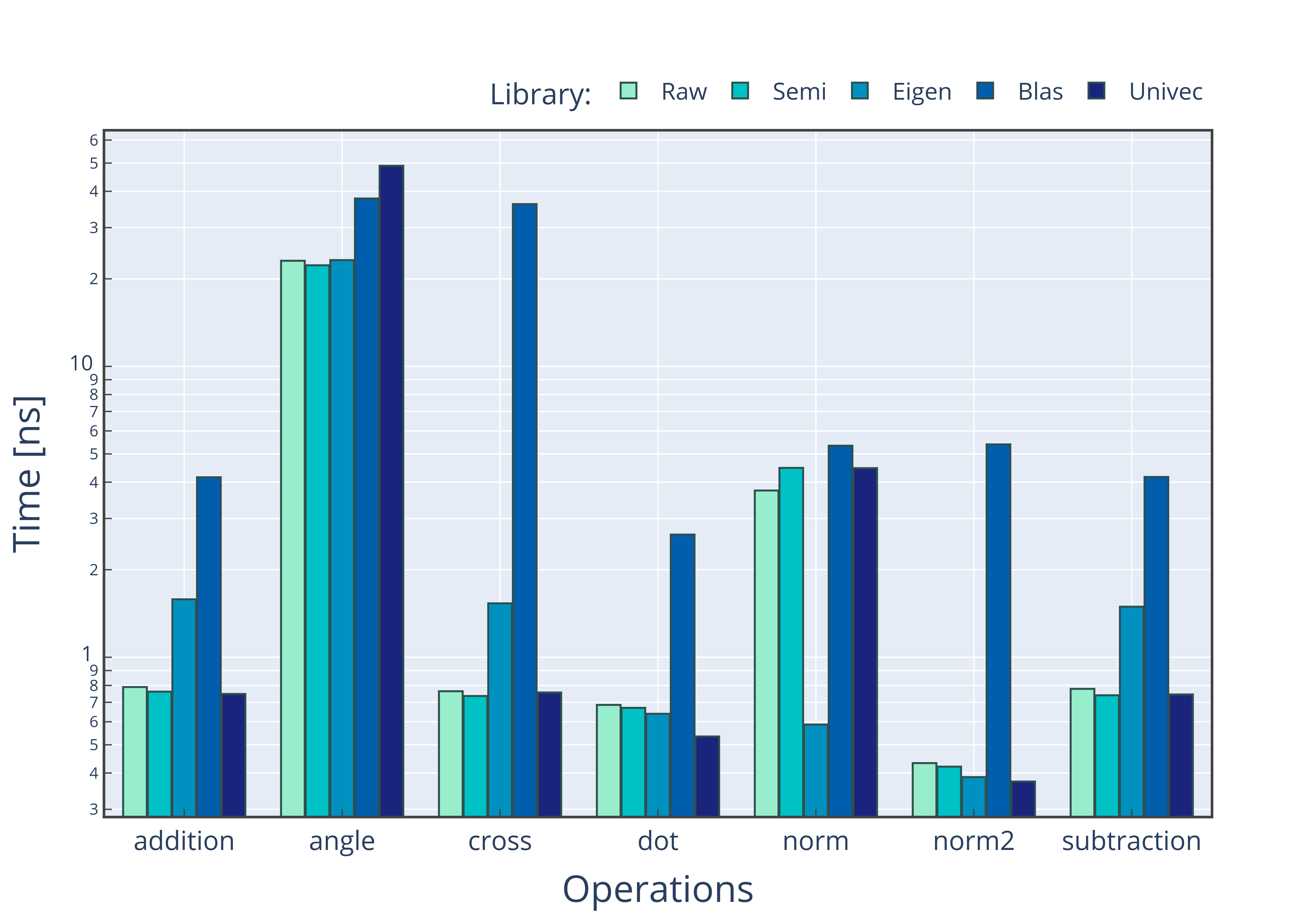}}

	\caption{Time needed to perform the specified vector operations, using five different code implementations. The left side refers to Linux setup, while the right side refer to macOS setup. In this plot, lower values mean better performance.}
	\label{fig:execution_performance_vector}
\end{figure*}

\begin{figure*}
	\subfigure[$2\times2$ matrices on Linux]{\includegraphics[width=0.49\textwidth]{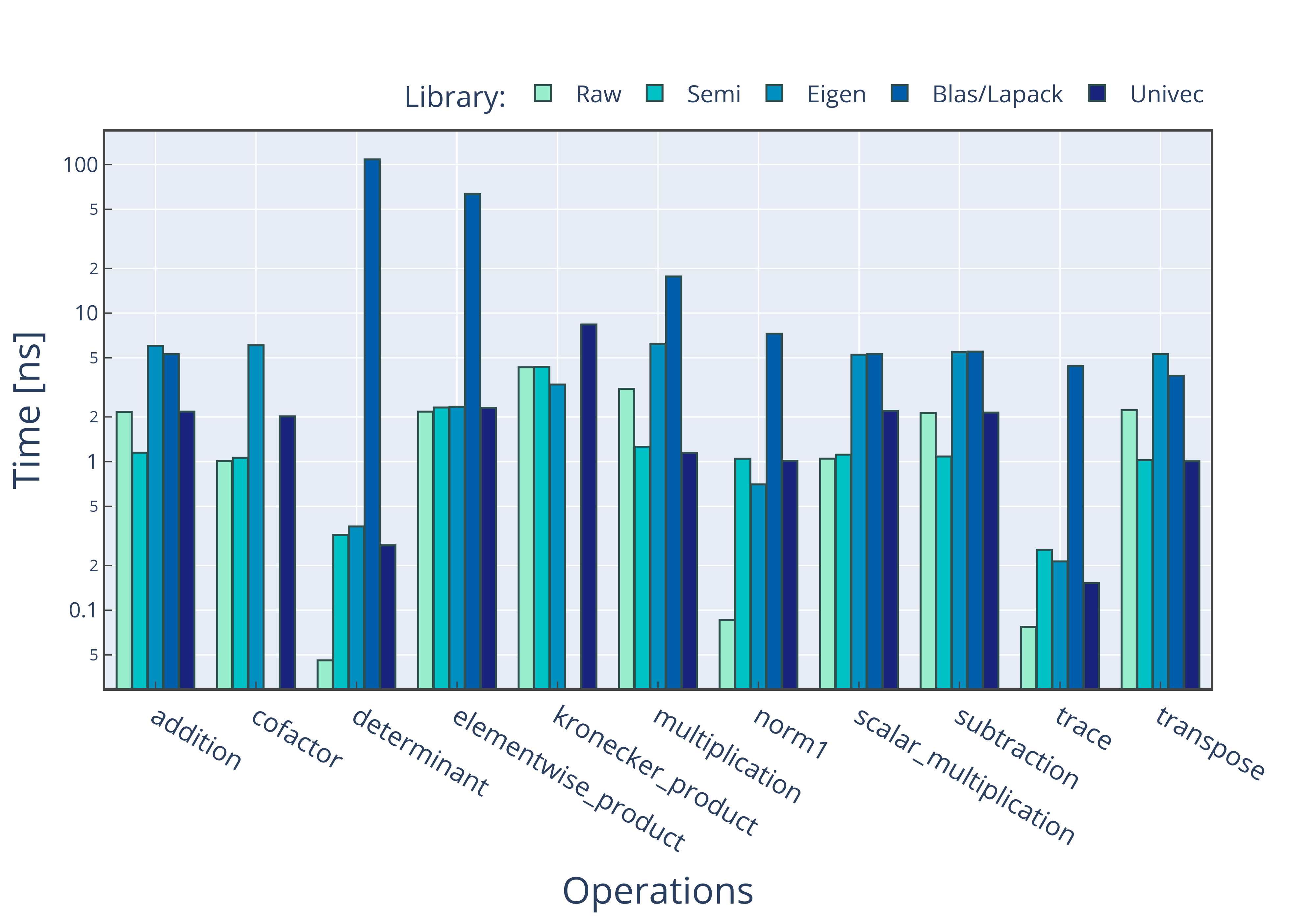}}
	\subfigure[$2\times2$ matrices on macOS]{\includegraphics[width=0.49\textwidth]{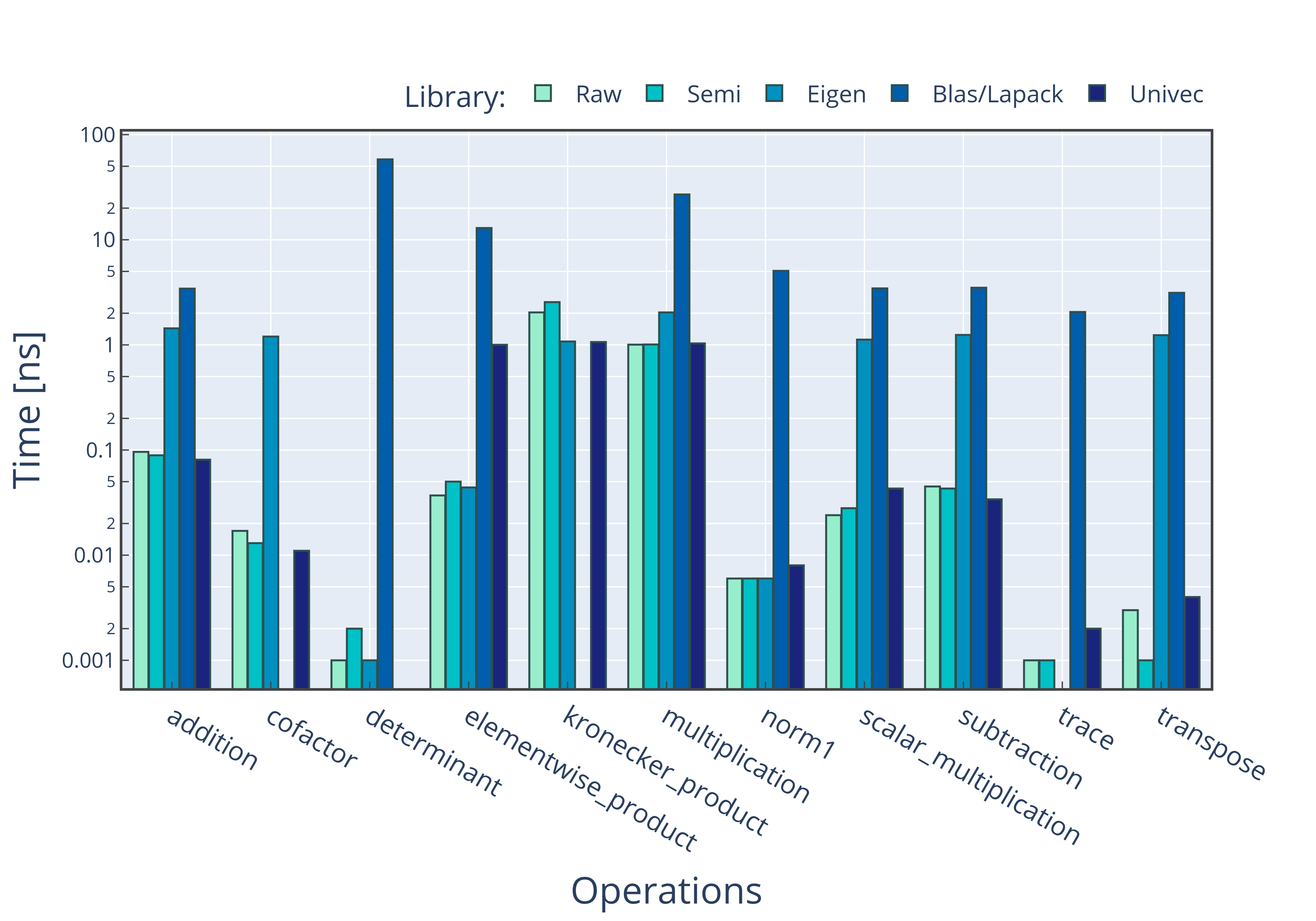}}
	
	\subfigure[$3\times3$ matrices on Linux]{\includegraphics[width=0.49\textwidth]{images/matrix_2x2_Linux.png}}
	\subfigure[$3\times3$ matrices on macOS]{\includegraphics[width=0.49\textwidth]{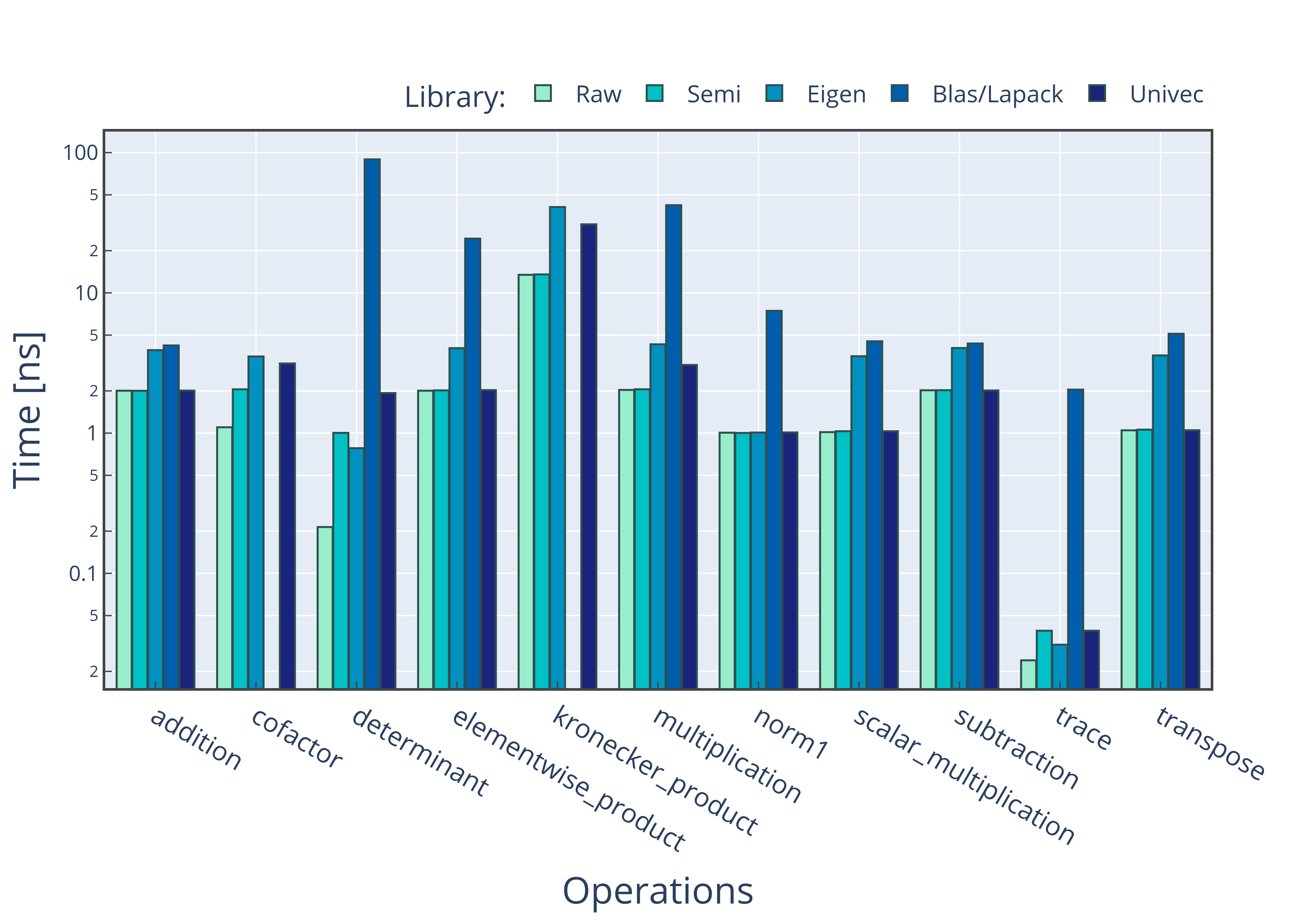}}
	
	\caption{Time needed to perform the specified matrix operations, using five different code implementations. The left side refers to Linux setup, while the right side refer to macOS setup. In this plot, lower values mean better performance.}
	\label{fig:execution_performance_matrix}
\end{figure*}

Performance comparison against existing solutions is a crucial aspect in the introduction of a new library. We compare \texttt{Univec} methods against four different approaches of performing the same operations:
\begin{itemize}
	\item \textbf{Raw}: using only primitive C++ types such as \texttt{double}.
	\item \textbf{Semi}: using \texttt{Boost::Units}\cite{boost_units} types and methods.
	\item \textbf{Eigen}\cite{eigenweb}: using \texttt{Eigen v3}\cite{eigenweb} types and methods.
	\item \textbf{BLAS}\cite{blasweb}/\textbf{LAPACK}\cite{lapackweb}: using \texttt{OpenBLAS}\cite{openblasweb} and \texttt{LAPACKE}\cite{lapackeweb} (for Linux) or the Accelerate framework\cite{accelerateweb} (for macOS) methods over primitive C++ types such as \texttt{double}.
	\item \textbf{Univec}: using \texttt{Univec} types and methods.
\end{itemize}
When the operation is too complex to implement (e.g. determinant of $N \times N$ matrix, where $N > 4$ for Raw and Semi approaches) or unavailable in a given library (e.g. cofactors of a $3 \times 3$ matrix for BLAS approach), the result is omitted. We compare a subset of operations, as shown in \cref{fig:execution_performance_vector,fig:execution_performance_matrix}. The procedure of performance testing is performed using this scheme:
\begin{enumerate}
	\item Input data are generated for Raw operation, using $n \cdot \num{10000}$ random values between \numrange{-1000}{1000}, with $n$ being the number of operands, and then stored in contiguous memory locations.
	\item Input data from Raw are copied to Semi, Eigen, Blas, and Univec. We use meter [\si{\meter}] unit for solutions with UoM validation.
	\item The timer is started.
	\item The operation is performed \num{10000} times, without multi-threading, and results are stored in a contiguous memory location.
	\item The timer is stopped, and the time spent for a single operation is calculated dividing the total time by \num{10000}.
	\item The results for each approach are compared for correctness against the result of Raw implementation.
	\item All these steps are repeated \num{1000} times in order to calculate the mean value and the stdev.
\end{enumerate}

\begin{figure*}
	\subfigure[Vector dot product on Linux]{
		\includegraphics[width=0.49\textwidth]{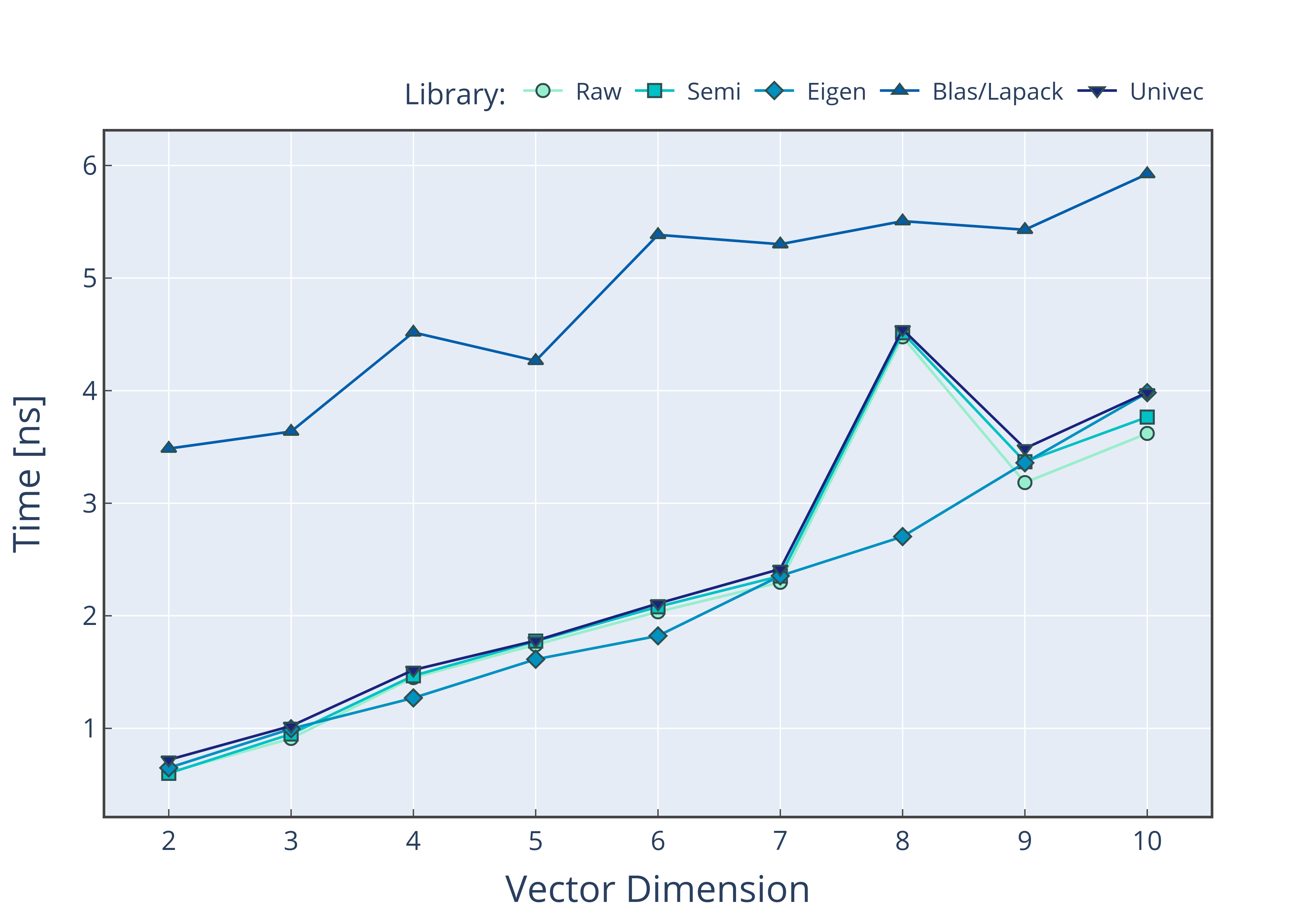}
		\label{fig:execution_scaling_vector_linux}
	}
	\subfigure[Vector dot product on macOS]{
		\includegraphics[width=0.49\textwidth]{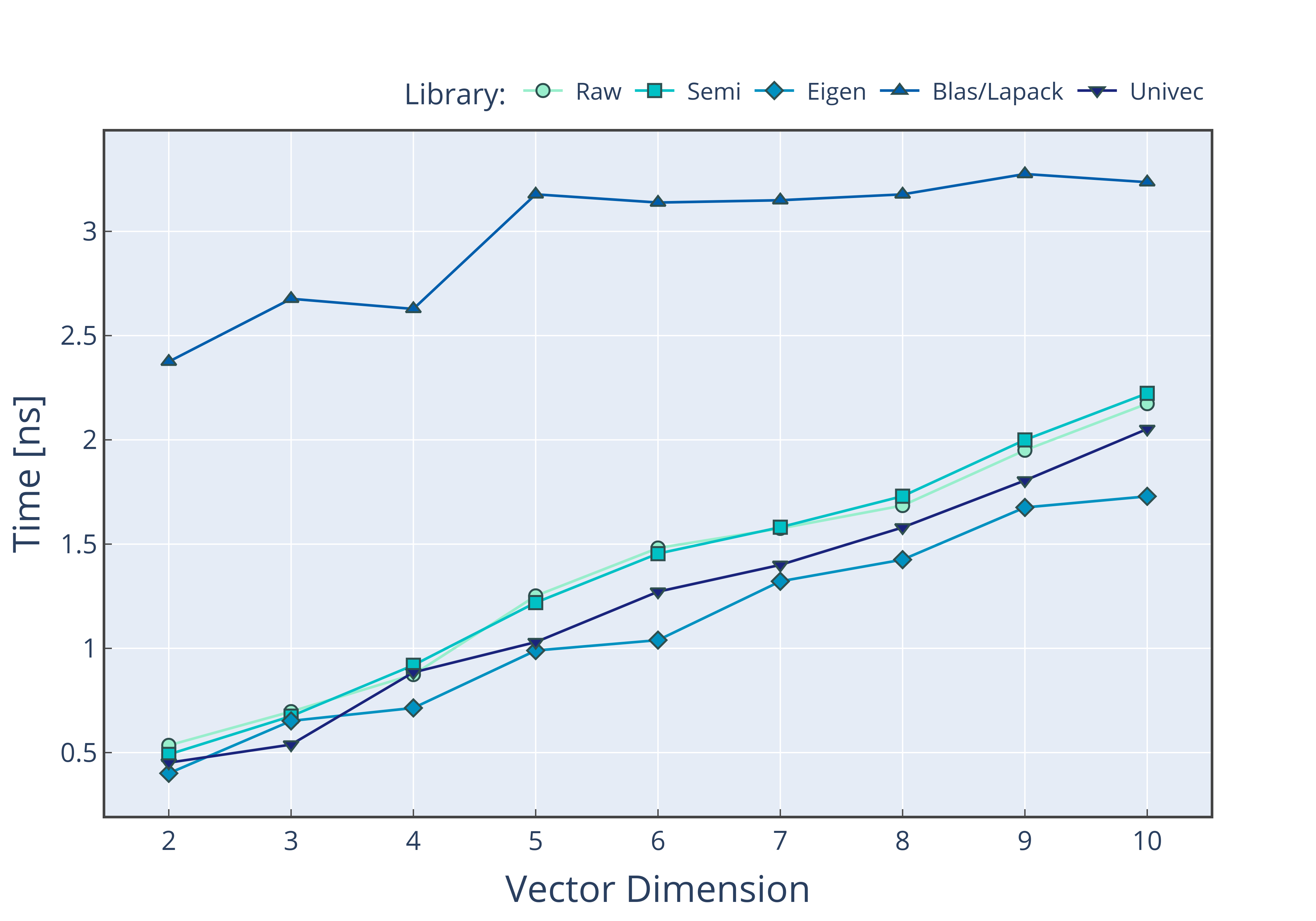}
		\label{fig:execution_scaling_vector_mac}
	}
	
	\caption{Time needed to perform the vector dot product for different vector sizes. The left side refers to Linux setup, while the right side refer to macOS setup. In this plot, lower values mean better performance.}
	\label{fig:execution_scaling_vector}
\end{figure*}

\begin{figure*}

	\subfigure[Matrix determinant on Linux]{
		\includegraphics[width=0.49\textwidth]{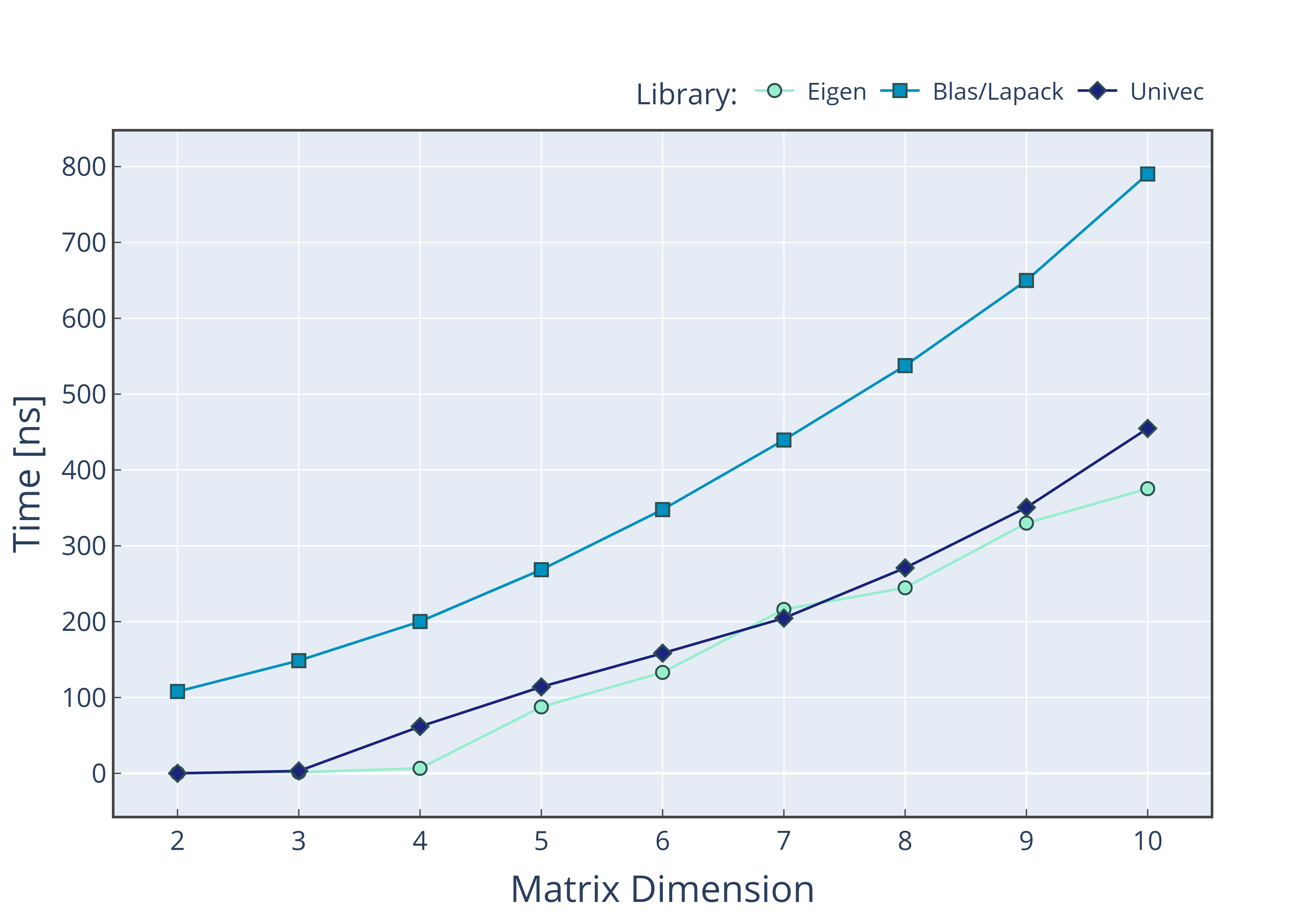}
		\label{fig:execution_scaling_matrix_linux}
	}
	\subfigure[Matrix determinant on macOS]{
		\includegraphics[width=0.49\textwidth]{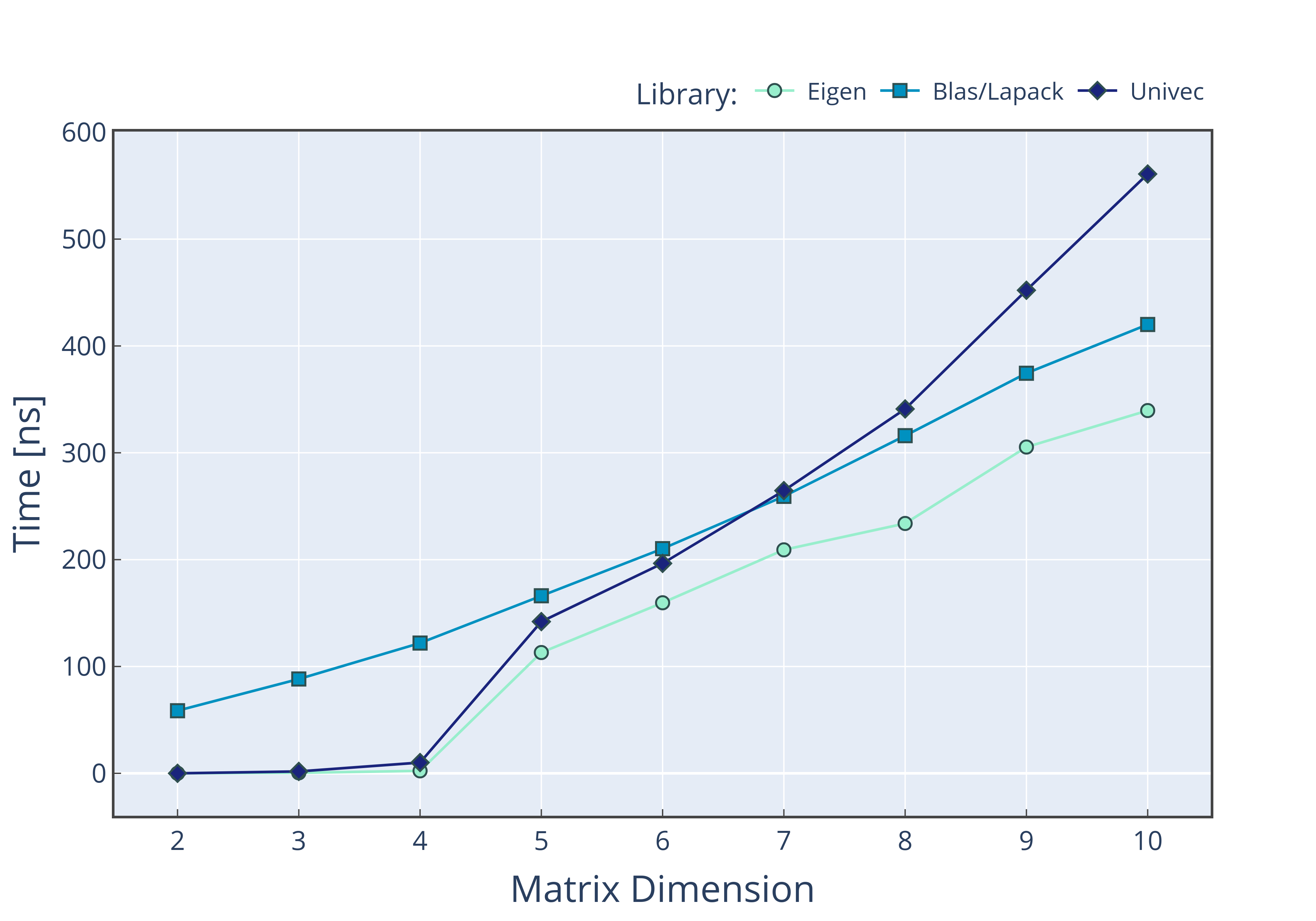}
		\label{fig:execution_scaling_matrix_mac}
	}
	
	\caption{Time needed to perform the matrix determinant for different matrix sizes. The left side refers to Linux setup, while the right side refer to macOS setup. In this plot, lower values mean better performance.}
	\label{fig:execution_scaling_matrix}
\end{figure*}

The source code for the performance analysis is available at the following locations:

\url{https://gitlab.com/edystan/univec-performance-test}

\url{https://gitlab.com/edystan/univec-scaling-test}

The first repository contains the performance tests for commonly used methods for both \texttt{Matrix} and \texttt{VectorC} classes, results being displayed in \cref{fig:execution_performance_vector,fig:execution_performance_matrix}. In contrast, the second one contains the performance analysis for only two methods (vector dot product and matrix determinant) but for vector dimensions between $2D$ and $10D$ and matrices sizes between $2\times2$ and $10\times10$. For each of the two performance projects, the test results are collected and saved in \texttt{.csv} files, one for each operating system, and available at the previously mentioned addresses. As well, results for these tests are shown in \cref{fig:execution_scaling_vector,fig:execution_scaling_matrix}.

The performance validation was performed on two laptops to simulate real scenarios in two operating systems. The configurations are the following:
\begin{enumerate}
	\item Lenovo V17 G2 ITL laptop with Intel Core i7-1165G7 at 2.80 GHz, 16 GB RAM. Ubuntu 22.04.2 LTS, Boost 1.74.0, Eigen 3.4.0, Univec 1.2, OpenBlas 3.10.0, Lapack 3.10.0. Compiled with GCC 11.2.0 with Release configuration.
	
	\item Apple MacBook Pro laptop with Apple M2 Max CPU at max. 3.49 GHz, 32 GB RAM. macOS Ventura 13.4.1, Boost 1.82.0, Eigen 3.4.0, Univec 1.2, Accelerate-1 (for Blas/Lapack). Compiled with Apple Clang version 14.0.3 with Release configuration.
\end{enumerate}

We can conclude that \texttt{Univec} performs in line with the other solutions, without any significant overhead, but with the addition of compile-time validation and a reasonable interface. In the performance analysis, we can see that Blas often shows poor performance, but we assume two factors cause this:
\begin{enumerate}
	\item The overhead of calling a Fortran function from C++ code (when applicable).
	\item The fact that we perform a BLAS call for each vector: we realize that BLAS can be used for some operations, but not all, in a SIMD (single instruction multiple data) fashion, which could mitigate the function calling overhead.
\end{enumerate}

In addition to this, we noticed an anomaly on the calculation of dot product for $8D$ vectors: as shown in \cref{fig:execution_scaling_vector_linux}, we can see that we have a performance hit on Linux/GCC platform, which seems to only appear at this specific dimension. We do not have a definitive explanation for this phenomenon, however we suspect that is a bug on the GCC optimization algorithm, because the same issue is not present in macOS/Clang environment, as show in \cref{fig:execution_scaling_vector_mac}.

Another anomaly we found in \cref{fig:execution_scaling_matrix_mac}, is that, for macOS, the timing of determinant do not scale well with the increasing of dimensionality. We believe that the causes of this behavior are:
\begin{enumerate}
 	\item The use the Gaussian elimination method for matrix bigger than $4\times4$, which is not the most efficient method available (BLAS uses the more efficient $PLU$ decomposition).
 	\item Some different optimization settings applied by the Apple Clang compiler.
\end{enumerate}

We plan to analyze and fix these optimization issues in the near future, in order to have performance in line with other implementations.

\section{Limitations and future plans}\label{sec:limits}

Our strongest limitation is that, currently, our solution only works with \texttt{Clang} of version higher or equal to \texttt{14}. This is caused by our decision to use the latest \texttt{C++20} features like \texttt{requires} and \texttt{constexpr} / \texttt{consteval}. Compatibility with \texttt{GCC} is limited  at this moment because we use the \texttt{\_\_declspec(property)} attribute to give access to the vector components. \texttt{Univec} was not tested under the \texttt{MSVC} or \texttt{icc} compiler, however we do not expect blocking issues as long as \texttt{C++20} is supported.

The use of \texttt{Boost.Units} as an underlying library for \texttt{Univec} has several non-negligible weak points that, unfortunately, seem to have no solution in the near future.
The first limitation is that this library is available only as part of a more comprehensive framework, \texttt{Boost}, which many software developers often see as a heavy dependency. The second limitation is that the current implementation is based on heavy use of Boost Metaprogramming Library (MPL) and template meta-programming techniques, which produce error messages challenging to understand for the non-experienced software developers.

In \cref{subsec:ex-particle}, we present an example of an elementary particle physics calculus implementation. However, the use of SI units is not encouraged and is not widespread in this field. We plan in a near future to add support to \texttt{Boost::Units}\cite{boost_units} for natural units, allowing more easy adoption of our solution by the HEP community.

There is a proposal to introduce the support for UoM-aware units directly in the standard language (see Mateusz Pusz's talk at CppCon 2019 \cite{PuszMaMan}). This approach will solve the problems presented above and will provide a more reliable solution with a less steep learning curve.
We plan to create another implementation which uses a standard library implementation when it will become available.
However, it will take several years until such implementation will find its way into the \texttt{C++} standard library.
Last, but not least, we plan to introduce in the near future a SIMD optimization (e.g. \texttt{SSE 2} or the newer \texttt{AVX-512}), which can increase the performance by an order of magnitude when working with vectors or matrices.

\section{Conclusions} \label{conclusions}

In this paper, we presented \texttt{Univec}, our solution for UoM-validation in software development, aiming to improve the use of vector and matrix calculations in \texttt{C++} development.
After a brief overview of the existing UoM solutions, we discussed their limitations and presented our solution, with several specific user-case scenarios.

The elaboration of \texttt{Univec} started as an internal project during the designing phase of our \texttt{Betaboltz} \cite{renda_betaboltz_2021} project. While developing this project, we realized the benefits of integrating the UoM analysis directly into our source code, virtually removing the most common causes of mistakes. Even if this solution does not remove all the sources of error, \texttt{Univec} allowed us to focus on the main development workflow, increasing confidence in our implementation.

We plan to also create to a native implementation when a UoM-aware implementation will be published in the \texttt{C++} standard library and we hope that similar solutions will be widely accepted in the development of advanced computational calculus for the scientific community.

\bibliography{refs}

\section*{Acknowledgements}

This study was supported by the Romanian Ministry of Research, Innovation and Digitization through the \texttt{PN23210104} and \texttt{ATLAS CERN-RO} grants.

\section*{Data availability}
All data generated or analysed during this study are included in this published article and its supplementary information files.

\section*{Author contributions statement}
\textbf{E.S.} Software, Validation, Writing - Original Draft, Visualization \\
\textbf{D.C.} Software, Validation, Writing - Review \& Editing \\
\textbf{M.R.} Conceptualization, Software, Writing - Review \& Editing, Supervision \\
\textbf{C.A.} Resources, Writing - Review \& Editing, Supervision, Project administration, Funding acquisition

\section*{Additional information}
The authors declare no competing interests.

\end{document}